\documentclass[aps,pra,print,groupedaddress,twocolumn,showpacs,10pt]{revtex4-1}
\usepackage{graphicx}
\usepackage{epsfig}
\usepackage{epstopdf}
\usepackage{subfigure}
\usepackage{appendix}
\usepackage{relsize}
\usepackage{amsmath,amssymb}
\usepackage{ulem}
\usepackage{color}
\usepackage[utf8x]{inputenc} 
\usepackage{etoolbox} 
\usepackage{lipsum} 
\usepackage[capitalize]{cleveref}
\begin{document}
\title{Single-photon transfer using levitated cavityless optomechanics}
\author{Pardeep Kumar}
\email[]{kxpsps@rit.edu}
\author{M. Bhattacharya}
\affiliation{School of Physics and Astronomy, Rochester Institute of Technology, 84 Lomb Memorial Drive, Rochester, NY 14623, USA}

\date{\today}

\begin{abstract}
We theoretically explore a quantum memory using a single nanoparticle levitated in an optical dipole trap and subjected to  feedback cooling. This protocol is realized by storing and retrieving a single photon quantum state from a mechanical mode in levitated cavityless optomechanics. We describe the effectiveness of the photon-phonon-photon transfer in terms of the fidelity, the Wigner function, and the zero-delay second-order autocorrelation function. For experimentally accessible parameters, our numerical results indicate robust conversion of the quantum states of the input signal photon to those of the retrieved photon. We also show that high fidelity single-photon wavelength conversion is possible in the system as long as intense control pulses shorter than the mechanical damping time are used. Our work opens up the possibility of using levitated optomechanical systems for applications of quantum information processing.
\end{abstract}


\maketitle

\section{Introduction}
Optomechanical systems provide a remarkable platform for controlling the interaction between photons and phonons at the quantum level  \cite{aspelmeyer2014,meystre2013}. Over the past few years, there has been a growing interest to harness these optomechanical interactions in quantum communication protocols \cite{cdong2015,felicetti2017}. In particular, cavity optomechanical oscillators have been explored as optical memory \cite{fiore2011,fiore2013} which allow light to be stored as a mechanical excitation and to be retrieved at any desired wavelength \cite{tian2010,dong2015,tian2015,stannigel2010,safavi2011}. Such protocols are useful in quantum and classical information processing since they permit the conversion of quantum states or traveling pulses between modes of vastly different frequencies \cite{tian2012,wang2012,clerk2012}. Interestingly, optomechanical light storage and retrieval has also been analyzed at a \textit{single}-photon level \cite{palomaki2013,galland2014}, thereby providing a promising platform for the transfer of quantum states \cite{filip2015}. This furnishes a testbed for verifying the quantum nature of photon-phonon-photon transfer \cite{filip2017,caprara2016,anderson2018}. Apart from this, optomechanical interactions have been exploited for ground-state cooling of the mechanical mode \cite{connell2010,chan2011}, sensing of the mechanical motion with imprecision below the standard quantum limit \cite{anetsberger2010}, strong coupling between optical and mechanical modes \cite{groblacher2009}, entanglement \cite{tpalomaki2013}, optomechanical squeezing \cite{pirkkalainen2015}, and optomechanically induced transparency \cite{agarwal2010,weis2010,zhou2013}. 
 
In spite of these significant applications, most experimental realizations in cavity-based optomechanical systems are hampered due to heating and decoherence produced by the mechanical clamping losses. Further, the use of cavities places  restrictions on the electromagnetic wavelengths, as they need to be resonant. A sensible solution to these limitations is provided by isolating the mechanical oscillator from its environment by means of levitation using optical \cite{neukirch2015,chang2009,yin2013} or magnetic \cite{romero2012,cirio2012} fields and using active feedback to substitute for the cavity. Such  levitated nanomechanical systems are approaching ground-state cooling \cite{li2011,gieseler2012,arita2013,jain2016,frimmer2016,rodenburg2016}, ultrasensitive applications \cite{moore2014,ranjit2015,frimmer2017,kumar2017,hebestreit2018,monteiro2018}, and preparation of quantum superposition states \cite{isart2010,isart2011}. 

Recently, it was proposed that levitated optomechanics could facilitate a favorable platform for the storage and retrieval of optical information at the multiphoton level \cite{kumar2018}.  In the present paper, we consider the optical memory protocol based on levitated cavityless optomechanics for the storage and retrieval of a \textit{single} photon, i.e., at the quantum level. Specifically, we investigate an effective protocol to store and retrieve quantum states \cite{wang2012} from the mechanical mode of the levitated optomechanical system. For this purpose, we consider the center-of-mass oscillations of a single nanoparticle levitated in an optical dipole trap subjected to nonlinear feedback \cite{lpneukirch2015}. The protocol follows a ``double swap'' scheme \cite{tian2010} by employing writing and readout pulses one mechanical frequency below the signal to effectively control the coupling between mechanical displacement and the signal. First, the writing and signal pulses arrive simultaneously to interact with the nanoparticle, thereby storing the quantum state of the signal in its mechanical mode. The system is then allowed to evolve freely for some time. Finally, the evolved state of the mechanical system is read at a later time utilizing a readout pulse.

We show that, under high vacuum, the Gaussian quantum states of a signal photon can be transferred with high fidelity \cite{isar2008} to the retrieved photon. However, the ambient conditions of pressure and temperature degrade the fidelity. We further characterize the effectiveness of the photon-phonon-photon transfer in terms of the Wigner function \cite{clerk2012} and the $g^{2}(0)$ function \cite{scully1997}, which we find remains one for the transfer of coherent states and becomes less than one for squeezed coherent states.  Moreover, we investigate the process in terms of the transmission of photon pulses \cite{tian2012}. We find the optimal pulse transmission under the condition of impedance matching \cite{dong2015,safavi2011}.  Our results suggest that a high pulse fidelity can be achieved by using input pulses of spectral width much narrower than the relevant transmission half-width \cite{tian2015}. This happens in the (multiphoton) strong-coupling limit where the transmission half-width remains independent of the mechanical decoherence, thereby allowing the efficient transfer of the input signal pulse. Thus we show that levitated optomechanical systems furnish a viable platform for quantum information processing tasks such as storage and retrieval. 
  
The structure of the remainder of the paper is as follows. In Sec. \ref{Sec2}, along with relevant equations, we describe the quantum memory protocol based on a single nanoparticle levitated in an optical dipole trap. Section \ref{Sec3} elucidates the effectiveness of the protocol in terms of the fidelity, the Wigner function, and the zero-delay second-order autocorrelation. A scattering matrix treatment of the pulse transmission is also presented. Finally, the concluding remarks of the paper are presented in Sec. \ref{Sec4}.
\vspace{-4mm}    
\section{Model}
\label{Sec2} 
\begin{figure} [ht]
   \begin{center}
   \begin{tabular}{c} 
   \includegraphics[height=7cm]{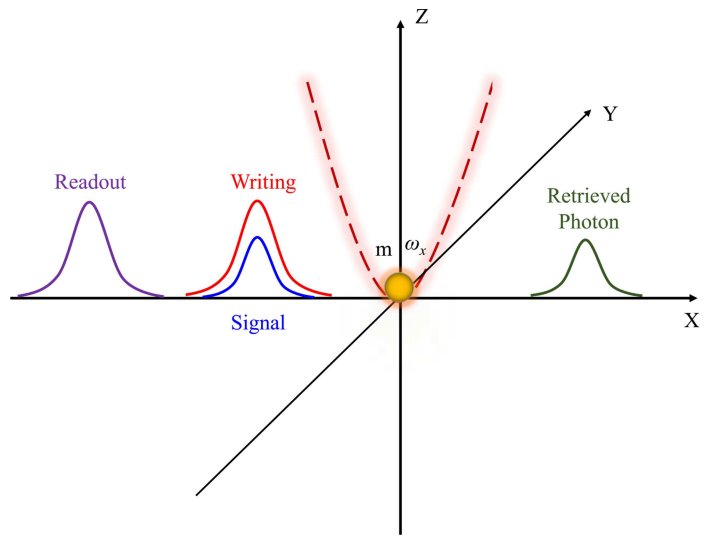}
	\end{tabular}
	\end{center}
   \caption[example] 
   { \label{fig1} Schematic of photon-phonon-photon transfer with an optically levitated nanoparticle of mass $m$ in a dipole trap and subjected to nonlinear feedback cooling \cite{rodenburg2016}. The particle oscillates in the optical dipole trap with frequency $\omega_{x}$ along the $x$ axis. A writing pulse along with the signal interact with the nanoparticle and causes the storage of the quantum states of the signal. The stored quantum state is retrieved at a later time by using a readout pulse.}
   \end{figure}

The system under consideration is a single dielectric nanoparticle of mass $m$ trapped in vacuum by a focused Gaussian beam. For small oscillation amplitudes, the three spatial modes of the mechanical oscillator are uncoupled and may be considered independently. Here, we consider the oscillations of the nanoparticle along $x$-direction such that its position is measured continuously by interferometric techniques \cite{gieseler2012,lpneukirch2015}. The monitored signal is then appropriately fed back \cite{rodenburg2016} to modulate the trap beam so as to cause additional damping of the nanoparticle, thereby giving rise to cooling, and also some backaction heating. Further, we use writing and readout pulses one mechanical frequency away from the signal to achieve the transfer of quantum states of a single photon through the levitated optomechanical system. The writing and signal pulses interact simultaneously with the nanoparticle, followed later by the readout pulse, as shown in Fig. \ref{fig1}. The presence of the writing pulse maneuvers the optomechanical coupling and facilitates the transfer of the quantum states of the signal to the mechanical oscillator. The stored photonic quantum states are retrieved at a later time once the readout pulse arrives.  

\subsection{Master Equation}
The quantum dynamics of the levitated nanoparticle are described by the following master equation \cite{rodenburg2016,ge2016}:
\begin{align}\label{eq1}
\dot{\rho}&=\frac{1}{i\hbar}\left[H,\rho\right]-\frac{\mathcal{A}_{t}}{2}\mathcal{D}\left[Q\right]\rho+\mathcal{L}_{sc}[\rho(t)]\nonumber\\
&-\frac{D_{p}}{2}\mathcal{D}\left[Q\right]\rho-\frac{D_{q}}{2}\mathcal{D}\left[P\right]\rho-i\frac{\gamma_{g}}{2}\left[Q,\left\{P,\rho\right\}\right]\nonumber\\
&-i\gamma_{f}\left[Q^{3},\left\{P,\rho\right\}\right]-\Gamma_{f}\mathcal{D}\left[Q^{3}\right]\rho\;,
\end{align}
where the first term on the right-hand side of Eq. (\ref{eq1}) represents the unitary evolution of the optomechanical system with $H=H_{0}+H_{int}$, where $H_{0}$ is the unperturbed Hamiltonian of the system and $H_{int}$ represents the optomechanical interaction, respectively.  The second term describes the positional decoherence of the nanoparticle due to the scattering of trap photons and $\mathcal{A}_{t}$ is the heating rate due to trap photon scattering \cite{rodenburg2016}. The third term represents the loss of photons from the optical field due to scattering from the nanoparticle and is given by $\mathcal{L}_{sc}[\rho]=-\mathcal{B}_{i}\left(\mathcal{D}[a]+\frac{2\omega_{i}^{2}\ell_{x}^{2}}{5c^{2}}\mathcal{D}[aQ]\right)\rho,~(i=w,s,r)$, where $\mathcal{B}_{i}$ represents the appropriate optical damping rate \cite{rodenburg2016}, $a~(a^{\dagger})$ is the annihilation (creation) operator of the optical field, $\omega_{i}$ is the optical frequency, and $\ell_{x}=\sqrt{\hbar/2m\omega_{x}}$ is the zero point fluctuation of the mechanical oscillator. Note that during writing process $\mathcal{B} = \mathcal{B}_{w} + \mathcal{B}_{s}$, while for readout process it becomes $\mathcal{B}_{r}$. The fourth and fifth terms describe the respective momentum and position diffusion of the nanoparticle due to collisions with background gas. The momentum (position) diffusion coefficient is $D_{p}=2\eta_{f}k_{B}T\ell_{x}^{2}/\hbar^{2} (D_{q}=\eta_{f}\hbar^{2}/24k_{B}Tm^{2}\ell_{x}^{2})$, where $T$ is the gas temperature, $k_{B}$ is Boltzmann's constant, $\eta_{f}=6\pi\mu R$ is the coefficient of friction, $\mu$ is the dynamic viscosity of the surrounding gas, and $R$ is the radius of nanoparticle. The sixth term represents gas damping at a rate $\gamma_{g}=\eta_{f}/2m$. The last two terms govern the nonlinear feedback damping and accompanying backaction, respectively. These terms are characterized by the respective coefficients  $\gamma_{f}=\chi^{2}\Phi \mathcal{G}$ and $\Gamma_{f}=\chi^{2}\Phi \mathcal{G}^{2}$. Here, $\chi, \Phi$, and $\mathcal{G}$ are the scaled optomechanical coupling, the average detected photon flux, and the feedback gain, respectively. The Lindblad superoperator in Eq. (\ref{eq1}) operates on the density matrix $\rho$ according to the rule $\mathcal{D}\left[\mathcal{K}\right]\rho=\left\{\mathcal{K}^{\dagger}\mathcal{K},\rho\right\}-2\mathcal{K}\rho \mathcal{K}^{\dagger}$, where $\mathcal{K}=Q,P,Q^{3},a$. Here, the mechanical position (momentum) quadrature is represented in dimensionless form as $Q=b^{\dagger}+b$ $\left(P=i(b^{\dagger}-b)\right)$. Also, $b~(b^{\dagger})$ is the annihilation (creation) operator of the mechanical oscillator.

\subsection{Assumptions} 
It is to be emphasized that the theoretical predictions of Eq. (\ref{eq1}) have been found to be in excellent agreement with experimental data \cite{rodenburg2016}. For the convenience of the reader, we summarize the main assumptions \cite{pflanzer2012} of our model as follows.\\
(1) The radius of the dielectric nanoparticle is assumed to be much smaller than the wavelength of the optical field.\\
(2) For amplitude of oscillations smaller than the beam waist and Rayleigh range, motion along three directions of oscillations is assumed to be uncoupled and can be treated independently.\\
(3) To derive Eq. (\ref{eq1}) \textit{Born-Markov approximation} is used since the coupling between the optical field and the background is assumed to be very weak and bath correlations decay very quickly. This approximation is valid under the condition if the bath correlation time is smaller than the relaxation time of the system \cite{breuer2002}. For our case, the bath correlation time at 4 K ($\tau_{B}\sim\frac{\hbar}{k_{B}T}\sim 10^{-11}$ s) is much smaller than the relaxation time ($\tau_{R}=\frac{1}{\Gamma}=1.5$ ms, where $\Gamma$ is the damping contributed by various relaxation processes, as explained below), thereby ensuring the validity of Born-Markov approximation. Further, the coupling between any systems or reservoirs is assumed to be small ($g\ll\Gamma,\mathcal{B}$, where $g$ is the single-photon optomechanical coupling). This ensures that the error in combining various master equations, e.g., optical and Brownian, is very small \cite{walls1970}.\\
(4) In the derivation of Eq. (\ref{eq1}), the terms oscillating at a high frequency have been neglected. This approximation is valid \cite{breuer2002} if the time for the intrinsic evolution of the system ($\tau_{S}=\frac{1}{\omega_{x}}=8~\mu$s) is smaller than the relaxation time ($\tau_{R}$). For our case, this condition yields $\Gamma<\omega_{x}$.\\  
(5) The Brownian motion term in Eq. (\ref{eq1}) characterizes the effect of collisions of the background gas with the nanoparticle. For the validity of this effect, a low density limit of the surrounding gas is assumed under the condition that the mean free path of the gas bath is smaller than the diameter of the nanoparticle \cite{breuer2002}.\\ 
(6) Further, to write the nonlinear feedback term in Eq. (\ref{eq1}), the Markovian limit is assumed where feedback is introduced rapidly as compared to any system time scale.\\  
(7) The possibility of interference between coherent and incoherent processes \cite{schwendimann1972} is assumed to be negligible for the system under consideration. The condition for this to occur is that the coherent coupling frequencies in the system be much larger than the inverse of the bath correlation times. However, in our case, the optomechanical coupling ($\sim$100 kHz) is much smaller than the thermal correlation frequency ($\sim \frac{1}{\tau_{B}}\sim10^{11}$ Hz).

\subsection{Quantum Langevin Equations}
Now the optomechanical interaction Hamiltonian is written as
\begin{align}\label{eq2}
H_{int}=\hbar ga^{\dagger}a\left(b^{\dagger}+b\right)\;,
\end{align}
where $g=V_{n}\frac{2\epsilon_{c}\omega_{s}\Delta\omega_{s}x_{0}}{\pi^{2}w_{0}^{2}c}\frac{\Delta x}{w_{0}^{2}}$ is the optomechanical coupling constant \cite{rodenburg2016,ge2016}. Here, $\epsilon_{c}$ is the effective relative permittivity of the dielectric, $V_{n}$ is the volume of the nanoparticle, $\omega_{s}$ is the frequency of the applied signal field, $\Delta\omega_{s}$ is the signal laser linewidth, and $w_{0}$ is the waist of the signal. We also have assumed that the focus of the signal field is shifted from that of the trap by a small amount $\Delta x$. In the presence of red-detuned coherent pulses \cite{kumar2018}, the linearized Hamiltonian in the rotating frame and in the interaction picture \cite{tian2012} is written as 
\begin{align}\label{eq3}
H_{int}=\hbar\Delta a_{s}^{\dagger}a_{s}+\hbar G_{i}\left(a_{s}^{\dagger}b+b^{\dagger}a_{s}\right)\;,
\end{align}
where $G_{i}=g\sqrt{n_{i}}$ is the effective optomechanical coupling rate, $n_{i}~(i=w,r)$ is the photon number of the writing (readout) field, $\Delta=\omega_{s}-\omega_{i}-\omega_{x}$ is the detuning,  $\omega_{i}~(i=w,r)$ is the frequency of the writing and readout field, and $a_{s}$ is the annihilation operator for the signal field. Note that, in Eq. (\ref{eq3}), we have used a mean-field approximation in which the intense control field can be treated classically and the optomechanical interaction is linearized with respect to the signal field \cite{fiore2011,linearization_reason}. Using the linearization process to describe optomechanical signal photon storage and retrieval, the master equation (\ref{eq1}) can be unraveled in terms of the following set of Langevin equations of motion: 
\begin{align}
\dot{a}_{s}&=-\left[i\Delta+\mathcal{B}\right]a_{s}-iG_{i}b+a^{s}_{in}\label{eq4}\;,\\ 
\dot{b}&=-\Gamma b-iG_{i}a_{s}+b_{in,\mathcal{T}}+b_{in,\mathcal{F}}\label{eq5}\;,
\end{align}
where $\Gamma = \gamma_{g} + \delta\Gamma$ is the mechanical damping \cite{damping_reason} and $\delta\Gamma=12\gamma_{f}\left(\langle N\rangle+\frac{1}{2}\right)$ is the nonlinear feedback damping. The stochastic terms have the correlations $\langle a^{s^{\dagger}}_{in}\left(t\right)a^{s}_{in}\left(t^{\prime}\right)\rangle=2\mathcal{B}\delta\left(t-t^{\prime}\right)$, $\langle b^{\dagger}_{in,T}\left(t\right)b_{in,T}\left(t^{\prime}\right)\rangle=2\gamma\delta\left(t-t^{\prime}\right)$, and $\langle b^{\dagger}_{in,\mathcal{F}}\left(t\right)b_{in,\mathcal{F}}\left(t^{\prime}\right)\rangle=2\mathcal{F}\delta\left(t-t^{\prime}\right)$. Here, $\gamma=2m\gamma_{g}k_{B}T_{eff}$, $T_{eff}$ is the effective temperature of the total background due to the combination of gas and optical scattering, and $\mathcal{F}=54m\hbar\omega_{x}\chi^{2}\Phi \mathcal{G}^{2}\left(2\langle N\rangle^{2}+2\langle N\rangle+1\right)$. The damping and noise terms due to nonlinear feedback depend on the mechanical state of the system \cite{rodenburg2016}. Note that, in the absence of damping and noise terms in Eqs. (\ref{eq4}) and \ref{eq5}), $\pi/2$ control pulses facilitate the complete mapping of motional state to optical state and vice versa thereby providing a realization of the optomechanical storage and retrieval \cite{fiore2011}. Further, it was recently shown that in the absence of backaction terms, Eqs. (\ref{eq4}) and \ref{eq5}) can be solved to demonstrate multiphoton storage using levitated cavityless optomechanics \cite{kumar2018}. Moreover, it was shown that such a cavityless protocol is compatible  with  wavelength conversion as well as for the retrieval of photons in the same direction as that of the incoming photons owing to the conservation of momentum.  Taking  these arguments into consideration, in this paper, we  include  the backaction terms in Eqs. (\ref{eq4},\ref{eq5}) to address the topic of light storage at a single-photon level. 
\section{Results}
\label{Sec3}
We investigate a protocol based on a single levitated nanoparticle to upload, store, and retrieve a single-photon quantum state from a mechanical mode. To describe this protocol, we employ Gaussian writing (readout) pulse $G_{w}=G_{w0}\exp\left[-\frac{\left(t-t_{w}\right)^{2}}{2t_{1s}^2}\right]~\Big(G_{r}=G_{r0}\exp\left[-\frac{\left(t-t_{r}\right)^{2}}{2t_{2s}^2}\right]\Big)$, such that in Eqs. (\ref{eq4}) and \ref{eq5}) we use $G_{i}=G_{w}+G_{r}$.
Here $G_{w0}~(G_{r0}), t_{w}~(t_{r})$ and $t_{1s}~(t_{2s})$ represent the respective amplitude, central time and the width of the writing (readout) pulse. 
\subsection{Optomechanical photon storage and retrieval}
The idea for the efficient photon-phonon-photon transfer is based on the storage and retrieval of a single photon using a single nanoparticle levitated in an optical dipole trap subjected to nonlinear feedback. In such a protocol, the quantum states of the signal photon and the mechanical mode can be swapped by red-detuned strong coherent pulses. That is, first the signal pulse is stored as a mechanical excitation by means of a writing pulse and, at a later time, the readout pulse results in the retrieval of stored optical information. Note that such an optomechanical photon storage and retrieval is characterized by the interaction Hamiltonian in Eq. (\ref{eq3}) and can be described by an anti-Stokes process \cite{tian2015}. During the writing process, the incoming signal photon is converted into a coherent phonon of the mechanical oscillator by means of $\omega_{s}-\omega_{w}=\omega_{m}$. On the other hand, the readout pulse at a later time is scattered to produce a retrieved photon by absorbing a phonon of the mechanical mode \cite{qu2012}, thanks to the up-conversion process $\omega_{re}=\omega_{r}+\omega_{m}$. Here $\omega_{re}$ is the frequency of the retrieved photon. 
\begin{figure} [ht]
   \begin{center}
   \begin{tabular}{c} 
   \includegraphics[height=6cm]{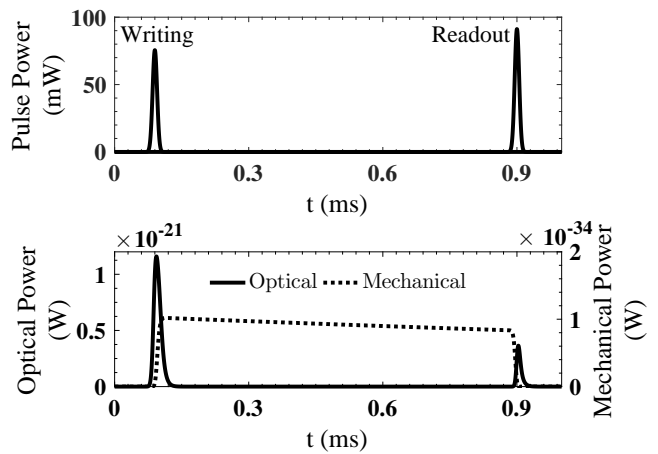}
	\end{tabular}
	\end{center}
   \caption{\label{fig2} Gaussian writing and readout pulses (top panel) and the calculated optical power of the storage and retrieval of the signal photon (solid line), along with the power of the stored mechanical oscillation (dotted line) as a function of time (bottom panel). The parameters used are $\omega_{x}$ = 124 kHz, $R$ = 50 nm, $m$ = 1.2$\times10^{-18}$ kg, $\epsilon_{c}$ = 1.133, signal wavelength ($\lambda_{s}$) = 780 nm, write wavelength ($\lambda_{w}$) = 1064 nm, readout wavelength ($\lambda_{r}$) = 1064 nm, central time of writing and signal pulse ($t_{w},t_{s}$) = 0.09 $m$s, central time of readout pulse ($t_{r}$) = 0.9 $m$s, writing and readout pulse widths ($t_{1s},t_{2s}$) = 7 $\mu$s, single photon signal optomechanical coupling ($g$) = 0.2 $m$Hz, effective writing (readout) optomechanical coupling ($G_{w0}~(G_{r0})$) = 79 kHz (86 kHz), $\Delta$ = 0, $\Delta x$ = 10 nm, $\ell_{x}$ = 19 $p$m,  $\gamma_{g}$ = 0.0289 Hz, $\delta\Gamma$ = 0.66 kHz, pressure ($P$) = 7 $\times$ $10^{-6}$ mbar, T = 4 K, $\mathcal{B}_{s}$ = 0.3 Hz, $\mathcal{B}_{w}$ = 0.04 Hz, $\mathcal{B}_{r}$ = 0.04 Hz, $\mathcal{A}_{t}$ = 27 kHz,  $\mathcal{A}_{w}$ = 10 kHz, $\mathcal{A}_{r}$ = 10 kHz, scaled optomechanical coupling $(\chi)$ = $1.5\times10^{-9}$, numerical aperture (NA) = 0.9 and nonlinear feedback gain ($\mathcal{G}$) = 20.}
   \end{figure} 
To illustrate optomechanical light storage and retrieval \cite{kumar2018}, we numerically solve Eqs. (\ref{eqa1}-\ref{eqa4}) in Appendix \ref{appendixA} [derived from Eqs.(4) and (5)] by using the Gaussian pulses as exhibited in the top panel of Fig. \ref{fig2}. The time dependence of the calculated power of the signal and retrieved photon, together with the power of the stored mechanical oscillation, is shown in the bottom panel of Fig. \ref{fig2}. For the storage process, a signal pulse which arrives simultaneously with writing pulse at $t_{s}$ = 0.09 ms is converted into mechanical excitation by means of a writing pulse. Subsequently, a readout pulse arrives at $t_{r}$ = 0.9 ms and converts the mechanical excitation back into the retrieved optical signal, as seen in Fig. \ref{fig2}. Such a protocol can be explored for the possibility of transfer of photonic states and we are encouraged in this endeavor by the high efficiency of photon retrieval during the readout process.
\subsection{Fidelity}
In the preceding section, we have described how the storage and retrieval of a single photon in levitated optomechanics sets a stage for the transfer of quantum states of a single photon. To characterize the quality of the retrieved photon, we calculate the fidelity defined by $F=\Big(\mbox{Tr}[\sqrt{\sqrt{\rho_{i}}\rho_{f}\sqrt{\rho_{i}}}]\Big)^{2}$, where $\rho_{i}$ is the initial density matrix of the signal photon and $\rho_{f}$ is the density matrix of the retrieved photon state. For Gaussian states, the fidelity can be calculated from the covariance matrices of the quadrature variables \cite{isar2008}. In order to do so, we assume the signal photon to be in a squeezed coherent state $|\alpha,r\rangle$, where, $\alpha$ is the coherent amplitude of the state and $r$ is the squeezing parameter. The initial mechanical state, on other hand, is assumed to be thermal. Using the covariance matrix,  as can be derived from Eqs. (\ref{eqb8}) and (\ref{eqb9}) in Appendix \ref{appendixB}, the fidelity can be written in the following form:
\begin{align}
F=\sqrt{\frac{2}{\mathcal{A}}}\exp\left[\frac{\zeta^{2}}{\mathcal{A}}\left(\mathcal{I}_{1}\mathcal{A}_{22}+\mathcal{I}_{2}\mathcal{A}_{11}\right)\right]\;,\label{eq6}
\end{align}
where $\zeta=1-e^{-\mathcal{B}_{r}t_{2s}-\Gamma t_{f}-\Gamma t_{1s}}$, $\mathcal{A}=\mathcal{A}_{11}\mathcal{A}_{22}$, $\mathcal{A}_{11}=e^{-2r}+V_{XX}$, $\mathcal{A}_{22}=e^{2r}+V_{YY}$, $\mathcal{I}_{1}=\left(\mbox{Re}(\alpha(0))\right)^{2}$,  $\mathcal{I}_{2}=\left(\mbox{Im}(\alpha(0))\right)^{2}$ and $t_{f}$ is the free evolution time. To describe the fidelity, we set  $G_{w}(t)=G_{w0}$ and $G_{r}(t)=G_{r0}$ and also consider $\pi/2$ writing and readout pulses. 
\begin{figure}[ht!]
\begin{center}
\begin{tabular}{ccc}
&\subfigure[]{\includegraphics[scale=0.305]{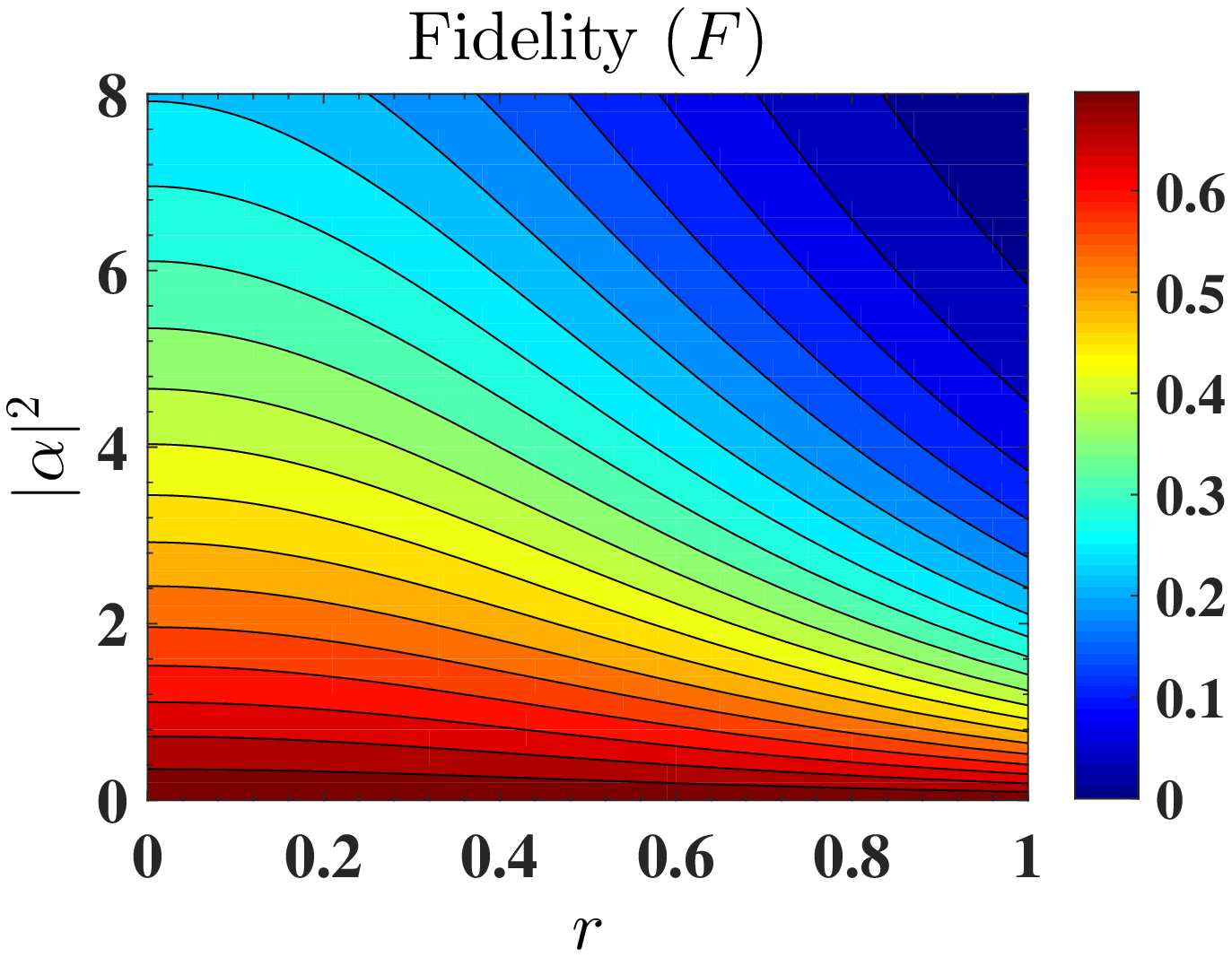}}  \subfigure[]{\includegraphics[scale=0.305]{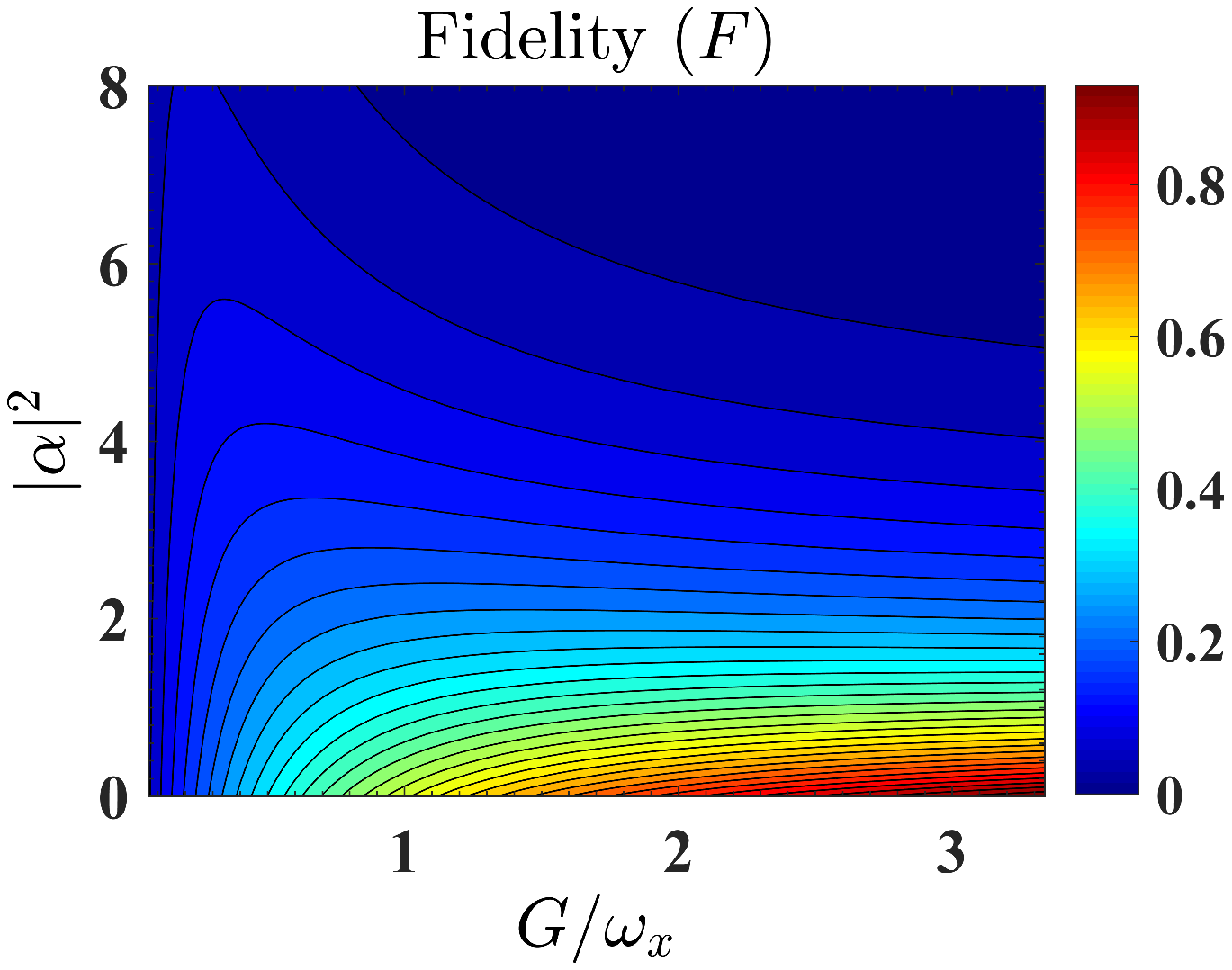}}&\\
&\subfigure[]{\includegraphics[scale=0.305]{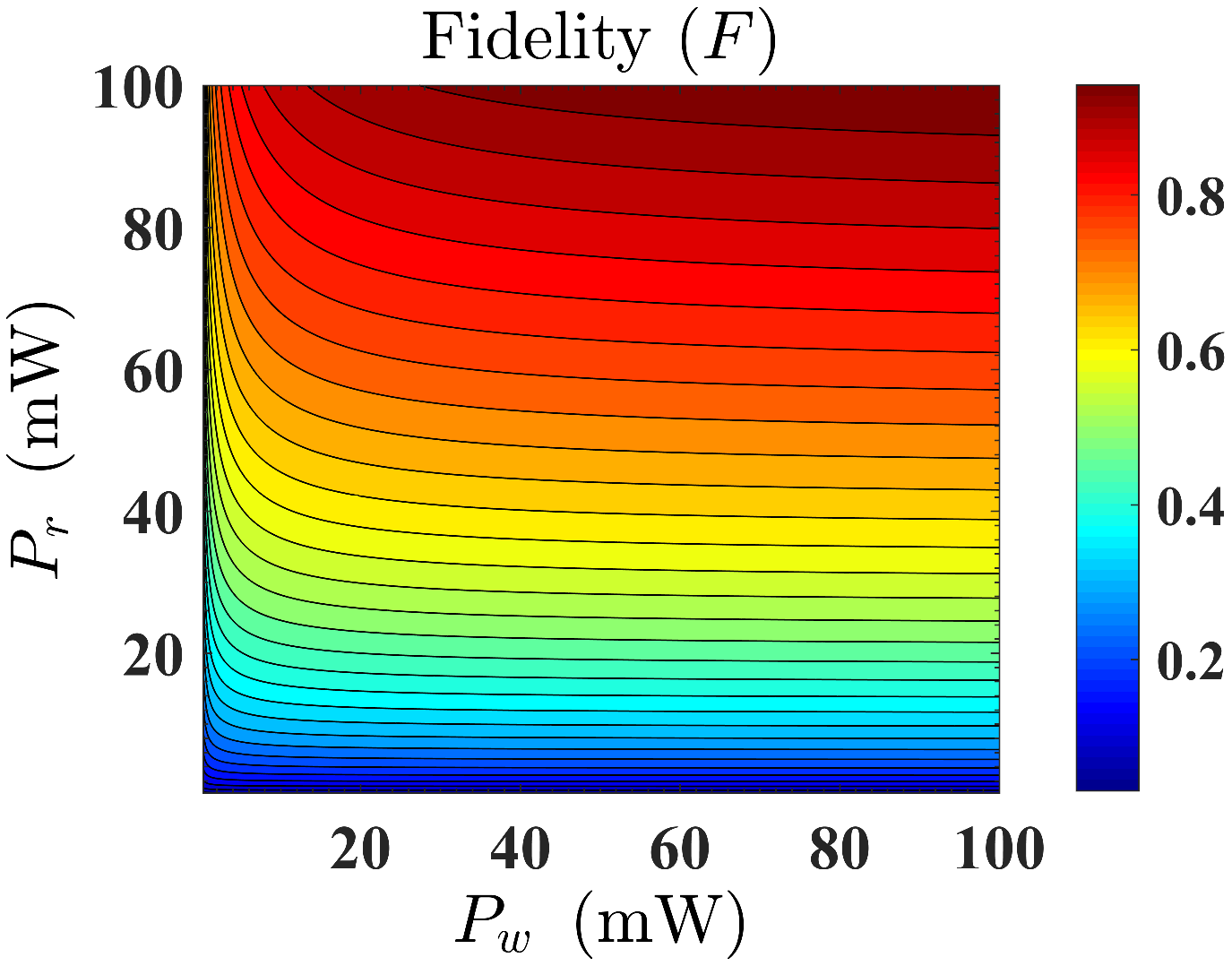}} & 
\end{tabular}
\end{center}
\caption{Fidelity vs (a) $r$ and $|\alpha|^{2}$, (b) $G/\omega_{x}$ and $|\alpha|^{2}$, and (c) power of writing [$P_{w}~(\mbox{mW})$] and readout pulses [$P_{r}~(\mbox{mW})$]. In plot (a) $t_{1s}$ = $19~\mu$s, $t_{2s}$ = $18~\mu $s, in plot (b) $G_{w0}$ = $G_{r0}$ = $G$, $r$ = 0.05, and in plot (c) $\alpha=\sqrt{0.3}$. Here, $t_{f}$ = 0.5 ms and other parameters are the same as in Fig. \ref{fig2}.}
\label{fig3}
\end{figure}

In Fig. \ref{fig3}(a), we study the dependence of the fidelity on the parameters of the initial squeezed coherent state. It is determined from Fig. \ref{fig3}(a) that a high fidelity can be achieved with small values of $\alpha$ and $r$. However, the fidelity deteriorates in the region of increasing parameters. The degradation of the fidelity with $\alpha$ is due to an increase in the average-amplitude decay of the retrieved photon state owing to the mechanical decoherence \cite{wang2012,tian2010}. On the other hand, the narrower features of the squeezed states in phase space cause the mechanical decoherence to become severe with increasing squeezing parameter $r$ thereby degrading the fidelity \cite{clerk2012}. Also, note that the high fidelity in Fig. \ref{fig3}(a) is attributed to $\pi/2$ pulses ($t_{1s}$ = 19 $\mu$s, $t_{2s}$ = 18 $\mu$s) which are relatively shorter than the mechanical decay time (30 ms) so as to provide immunity to the levitated optomechanical system against decoherence.  Further, the double swap scheme enables the Gaussian states to be transferred efficiently if the optomechanical coupling is stronger than mechanical decoherence rate ($G_{i}>\Gamma$). This is shown in Fig. \ref{fig3}(b), where a strong effective optomechanical coupling results in the high transfer fidelity. Such a strong optomechanical coupling can be achieved by employing intense writing and readout pulses so as to achieve high transfer fidelity, as exhibited in Fig. \ref{fig3}(c).     
\begin{figure}[ht!]
\begin{center}
\begin{tabular}{ccc}
\subfigure[]{\includegraphics[scale=0.305]{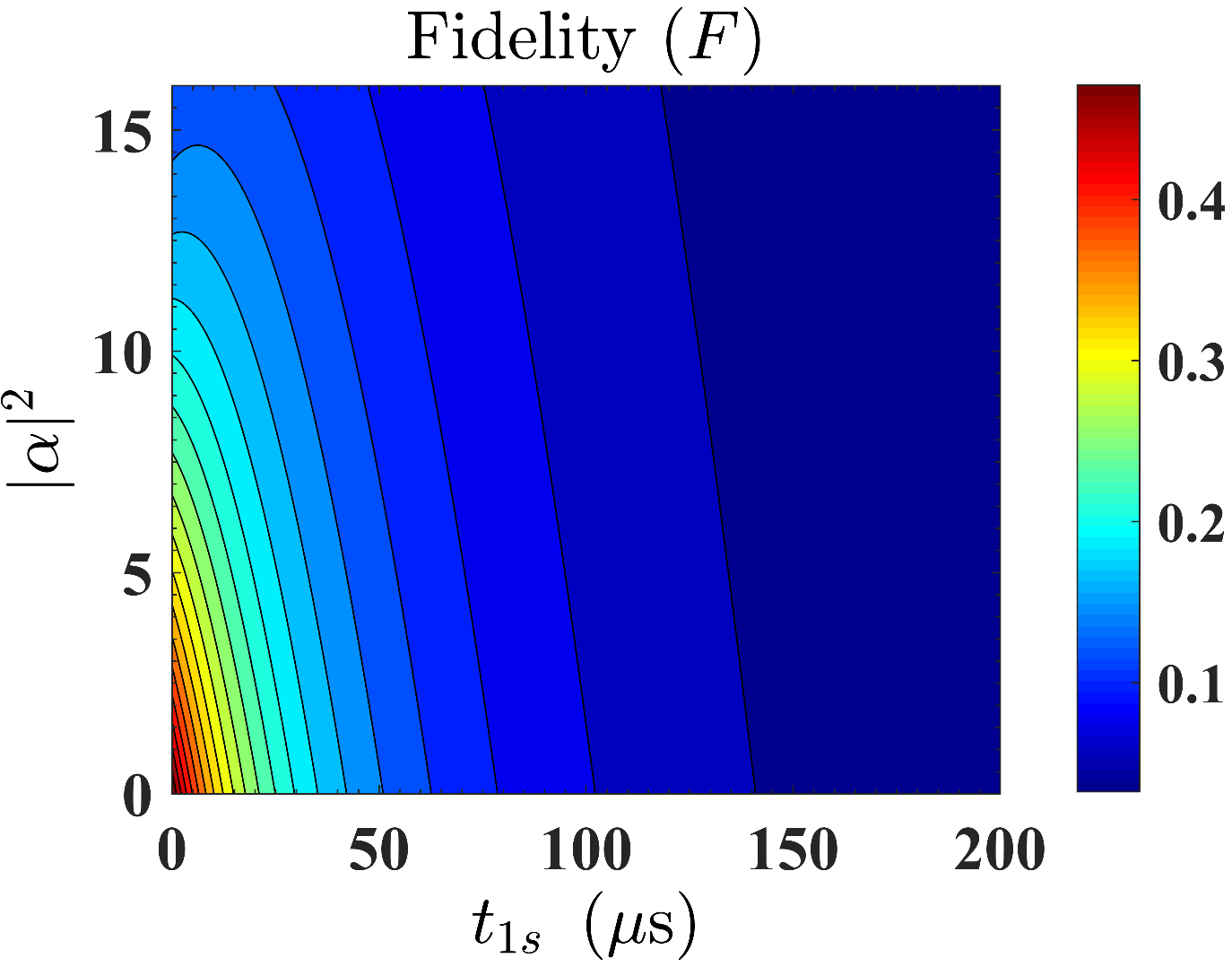}} &  \subfigure[]{\includegraphics[scale=0.305]{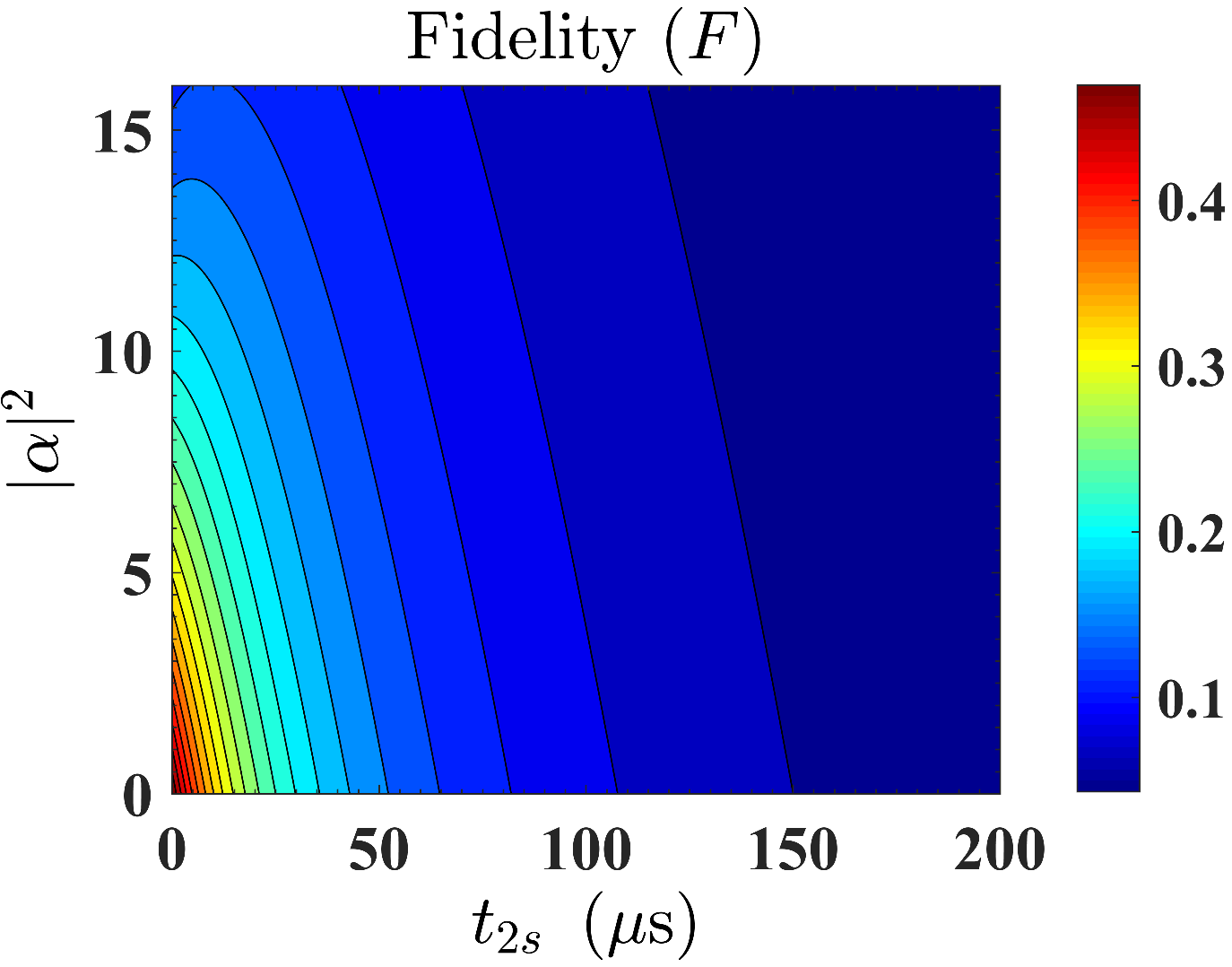}}
\end{tabular}
\end{center}
\caption{(a) Fidelity vs $t_{1s}~(\mu s)$ and $|\alpha|^{2}$. (b) Fidelity vs $t_{2s}~(\mu s)$, and $|\alpha|^{2}$. In plot (a) $t_{2s}$ = 7 $\mu$s and in plot (b) $t_{1s}$ = 7 $\mu$s and other parameters are the same as in Fig. \ref{fig3}.}
\label{fig4}
\end{figure}
 
To delineate the above results, we have used $\pi/2$-area writing and readout pulses. However, it is still possible to efficiently transfer the quantum states of a signal photon even if the pulse area exceeds $\pi/2$, thanks to the absence of clamping losses in levitated optomechanics. Nevertheless, the process involves writing and readout pulses of duration smaller than the mechanical decoherence time. Under this situation, the system remains immune to the decoherence, thereby providing a high fidelity in our protocol, as illustrated in Fig. \ref{fig4}(a) and (b). Despite the presence of the short span pulses, the fidelity again degrades with the amplitude of the coherent state of the signal photon for the same reason as explained above. 
\begin{figure}[ht!]
\begin{center}
\begin{tabular}{cc}
\includegraphics[scale=0.31]{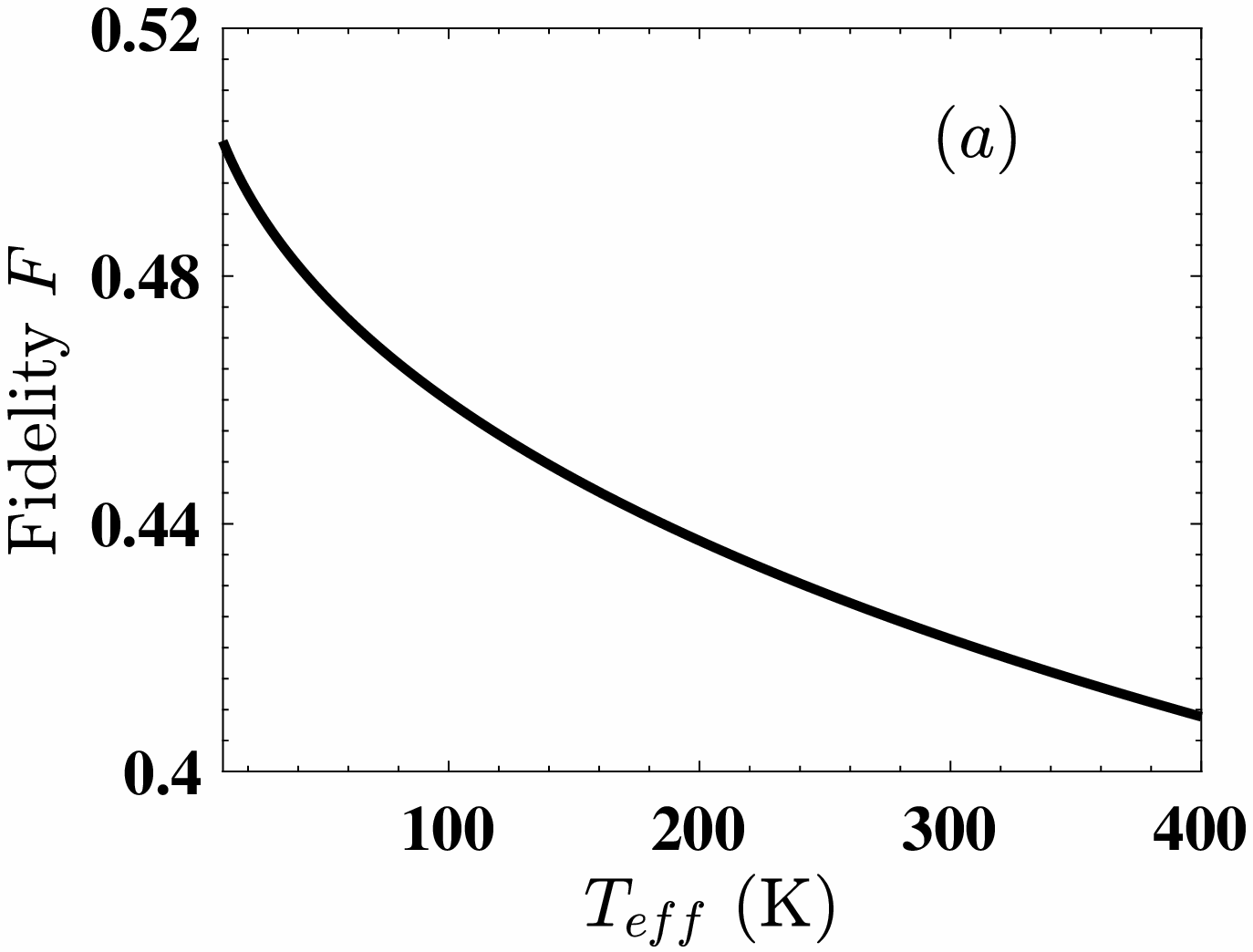} & \includegraphics[scale=0.31]{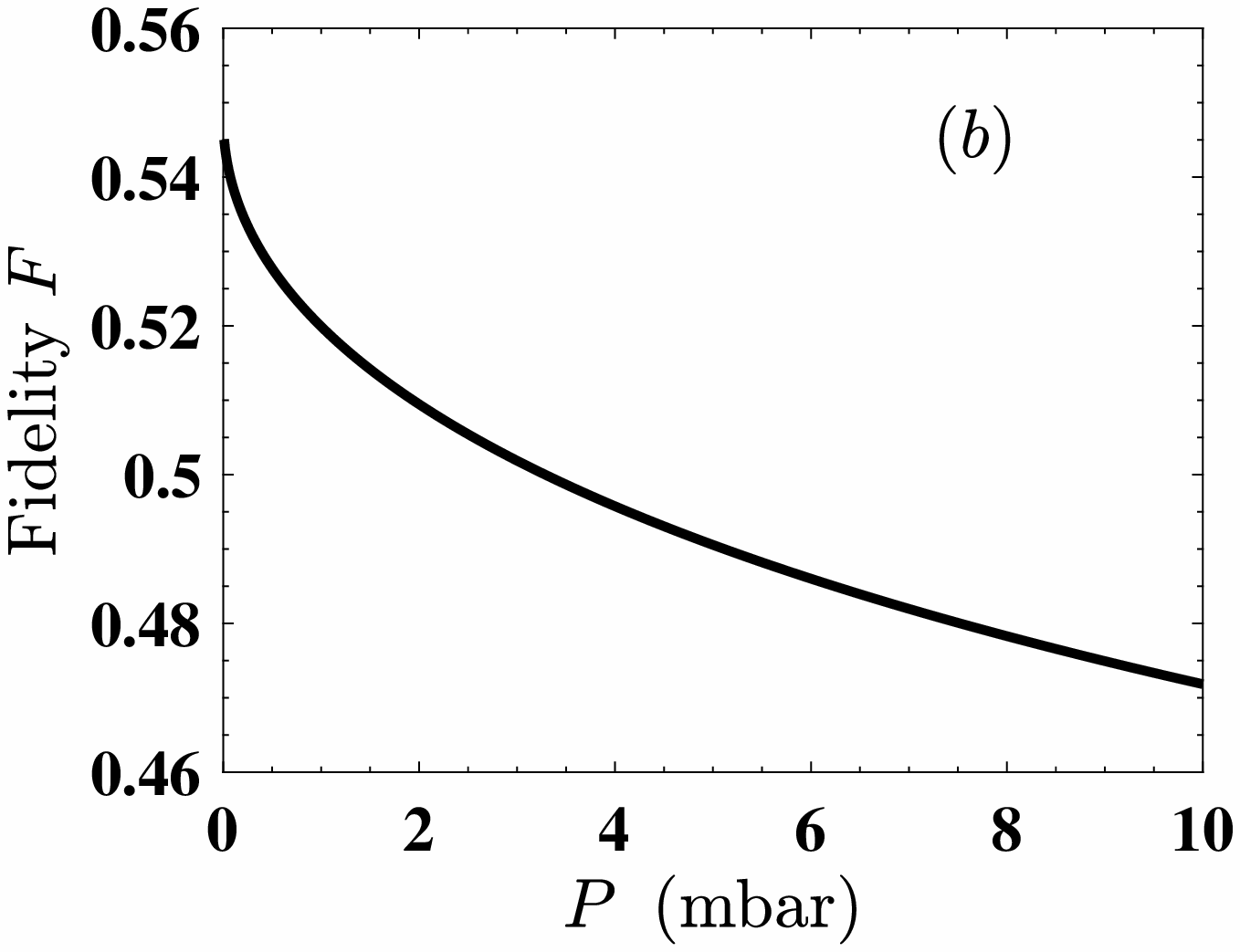}
\end{tabular}
\end{center}
\caption{Fidelity versus (a) temperature (K), and (b) pressure (mbar). Here, $t_{1s}=t_{2s}=$ 7 $\mu$s and the rest of the parameters are the same as in Fig. \ref{fig3}.}
\label{fig5}
\end{figure}

So far the preceding results are based on the conditions of low pressure ($\leq 10^{-5}$ mbar). Under this low-pressure regime, the preparation of oscillator near the ground state leads to an efficient retrieval of a photon from the mechanical oscillator, thereby providing a high value of fidelity as depicted in Fig. \ref{fig5}(a). However, in the high-pressure regime, the fidelity degrades due to an increase in the gas damping as exhibited in Fig. \ref{fig5}(b).

\subsection{Wigner function of the retrieved photon}
Above we have described that strong writing and readout pulses of short duration are suitable for the efficient transfer of Gaussian states of the signal. To further quantify the effectiveness of the photon-phonon-photon transfer, we write the following form of the Wigner function \cite{clerk2012} of the retrieved photon state by using covariance matrix ($V$) [derived from Eqs. (\ref{eqb8}) and (\ref{eqb9}) in Appendix \ref{appendixB}]: 
\begin{align}
W&=\frac{1}{2\pi \sqrt{\mathcal{V}}}\exp\left[-\frac{\xi^{2}}{2\mathcal{V}}\left(\mathcal{I}_{1}V_{YY}+\mathcal{I}_{2}V_{XX}\right)\right]\label{eq7}\;,
\end{align}
where we have introduced
\begin{align}
\xi&=\frac{\eta_{r}}{4}\Big[(\eta_{w}-\eta_{f})\cos(\theta_{+})+(\eta_{w}+\eta_{f})\cos(\theta_{-})\Big]\label{eq8}\;.
\end{align} 
Here $\eta_{w} = \exp{(-\mathcal{B}t_{1s})}$,  $\eta_{f} = \exp{(-\Gamma(t_{f}+t_{1s}))}$,  $\eta_{r} = \exp{(-\mathcal{B}_{r}t_{2s})}$, $\theta_{\pm} = G_{w} t_{1s}\pm G_{r}t_{2s}$ and  $\mathcal{V}=V_{XX}V_{YY}$. 
\begin{figure}[ht!]
\begin{center}
\begin{tabular}{cc}
\subfigure[]{\includegraphics[scale=0.305]{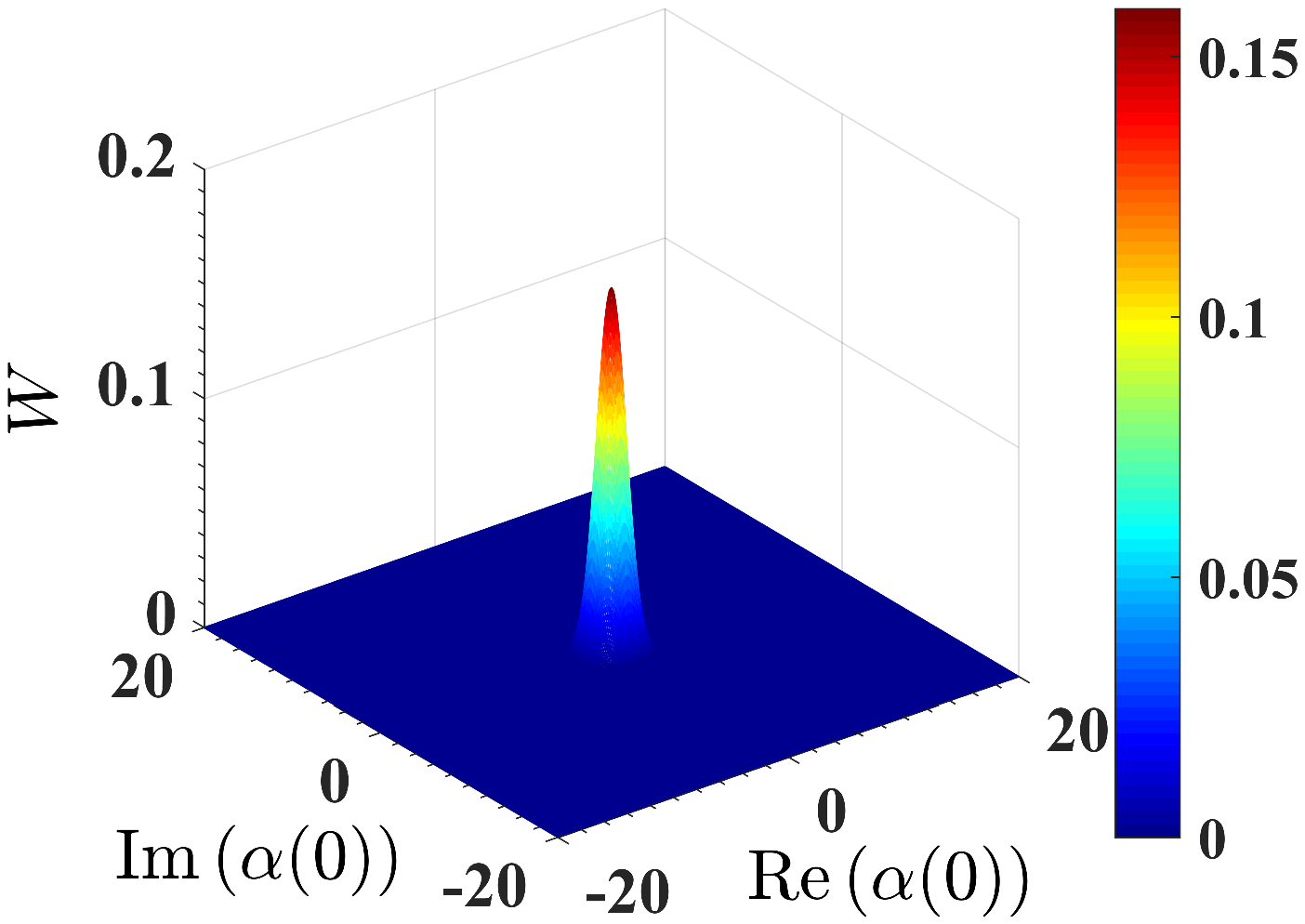}}& \subfigure[]{\includegraphics[scale=0.305]{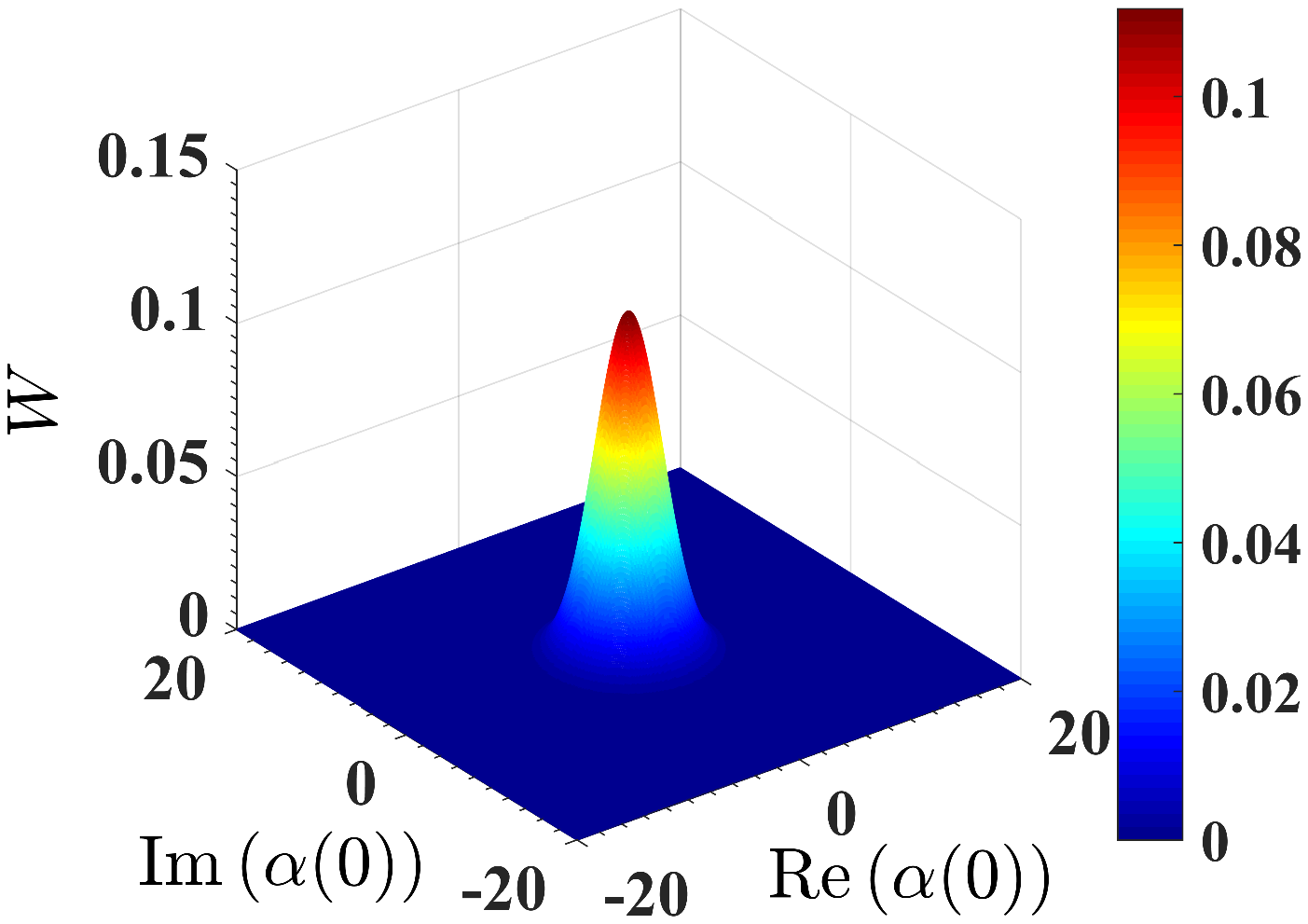}}\\
\subfigure[]{\includegraphics[scale=0.305]{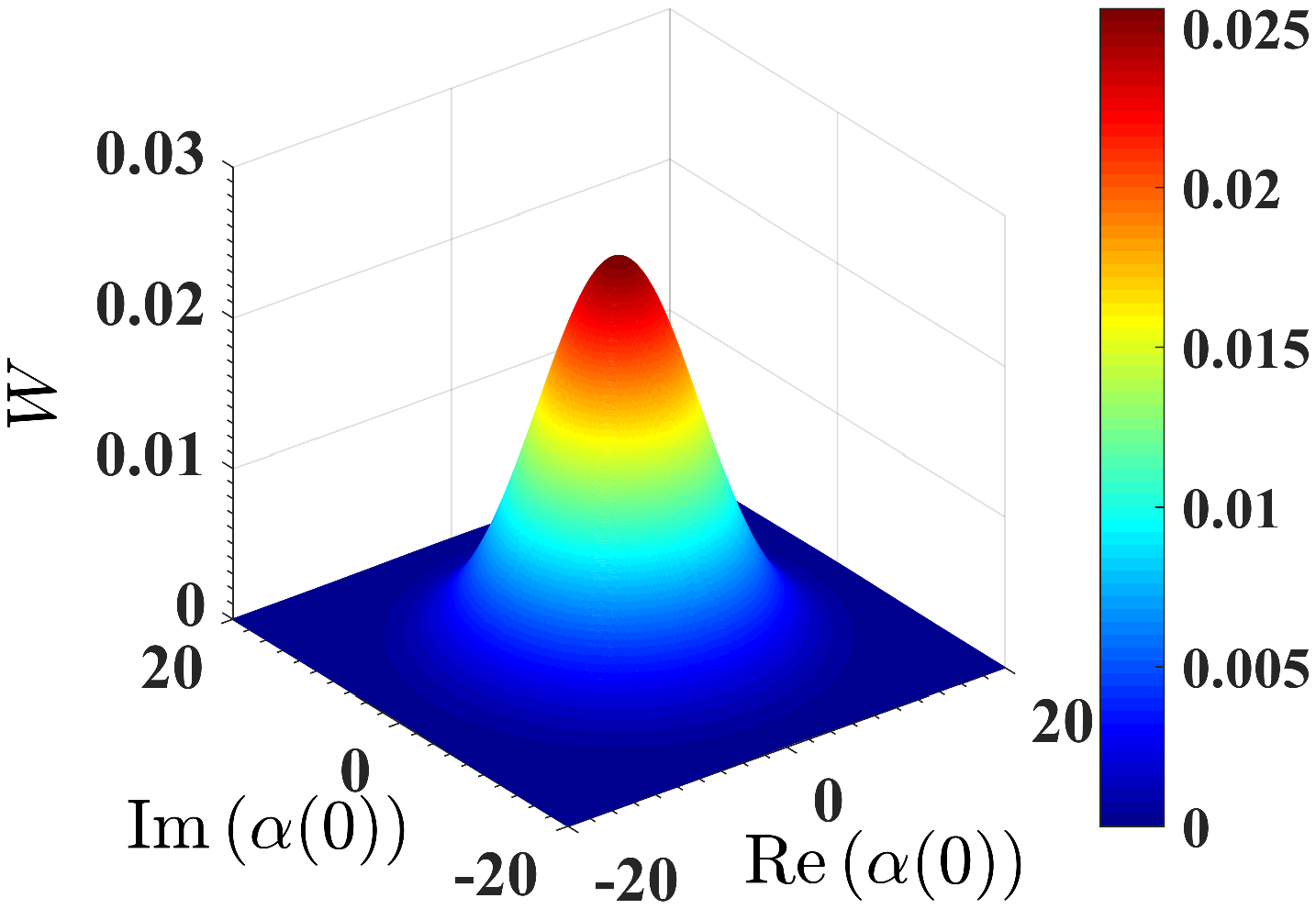}}& \subfigure[]{\includegraphics[scale=0.305]{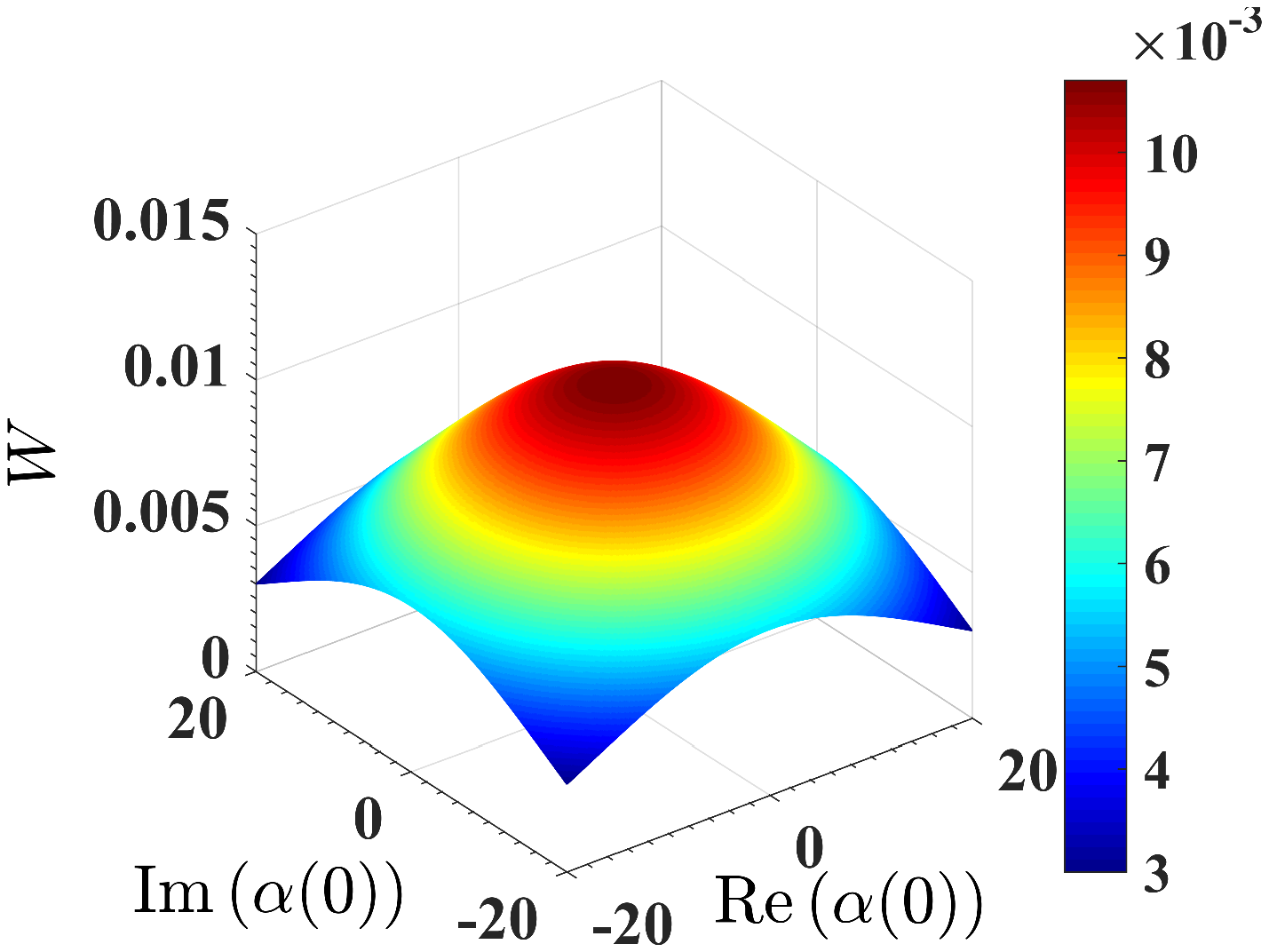}}  
\end{tabular}
\end{center}
\caption{(a) Wigner function of the input Gaussian state of the signal. The Wigner function of the retrieved photon state for (b) $t_{1s}$ = $t_{2s}$ = 1$~\mu s$, (c) $t_{1s}$ = $t_{2s}$ = 50$~\mu s$, and (d) $t_{1s}$ = $t_{2s}$ = 100$~\mu s$. Other parameters are the same as in Fig. \ref{fig3}.}
\label{fig6}
\end{figure}

Degradation due to mechanical decoherence limits the transfer fidelity, thereby causing the distortion in the shape of the Wigner function of the retrieved photon. This is shown in Fig. \ref{fig6}. For pulses of very short span, for instance, $t_{1s}$ = $t_{2s}$ = 1$~\mu s$ in Fig. \ref{fig6}(b), the Wigner function of the retrieved photon state remains Gaussian similar to that of the input signal state [see Fig. \ref{fig6}(a)]. However, the larger temporal width of the pulses causes mechanical decoherence to intervene in the process to produce distortion in the Wigner function as depicted in Figs. \ref{fig6}(c) and (d). Besides this, we have also figured out that in the presence of $\pi/2$ pulses and for $G_{i}>\Gamma$, the Wigner function of the retrieved photon remains relatively the same as that of input signal photon state. Thus the Wigner function provides a good measure of the effectiveness of the protocol.
\subsection{$g^{2}(0)$ function}
\begin{figure}[ht!]
\begin{center}
\begin{tabular}{cc}
\subfigure[]{\includegraphics[scale=0.295]{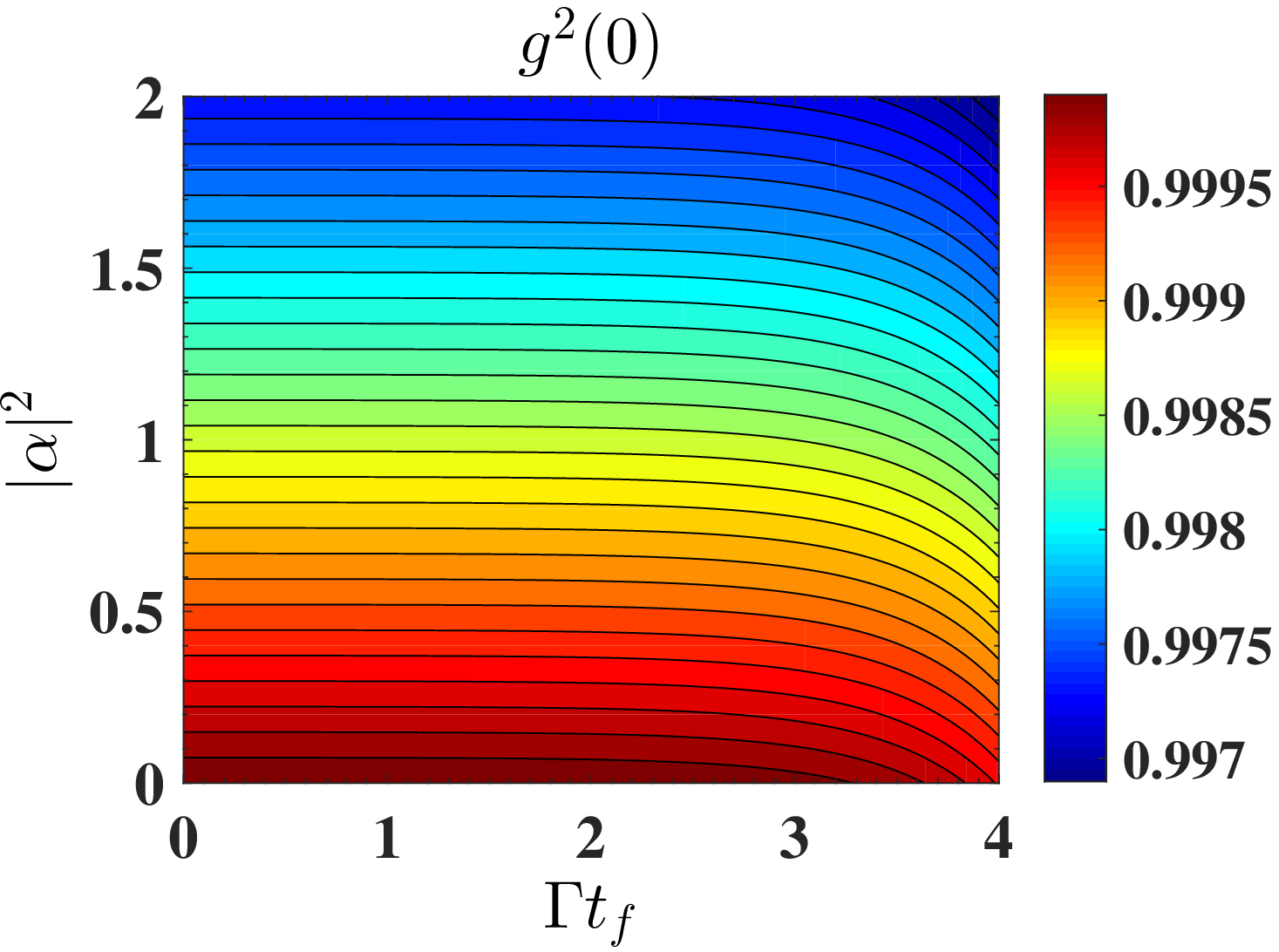}} & \subfigure[]{\includegraphics[scale=0.295]{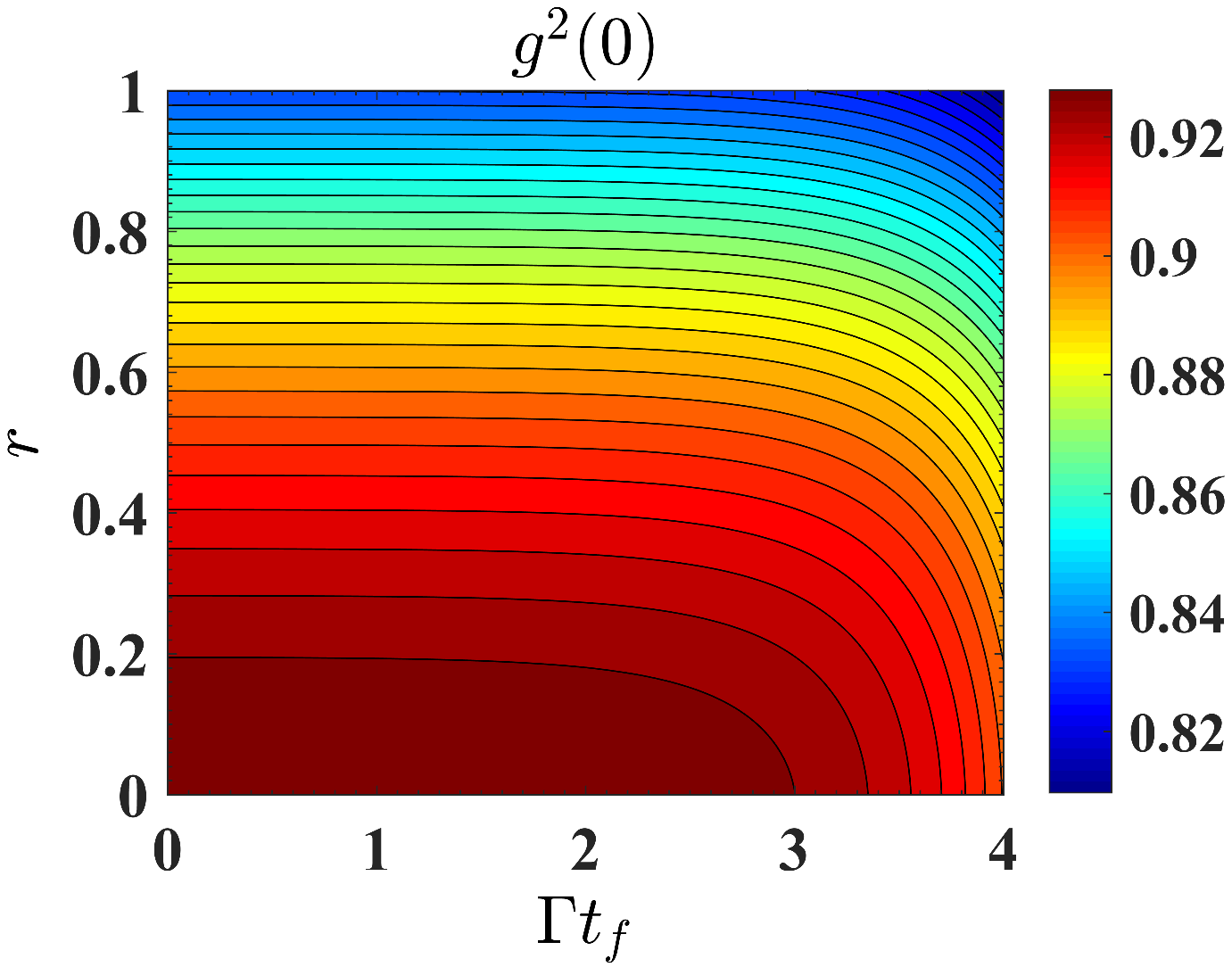}}
\end{tabular}
\end{center}
\caption{$g^{2}(0)$ function vs (a) $|\alpha|^{2}$ and $t_{f}$ and (b) $r$ and $t_{f}$. In plot (a) $r$ = 0, while in plot (b) $\alpha$ = $\sqrt{0.3}$ and rest of the parameters are same as in Fig. \ref{fig3}.}
\label{fig7}
\end{figure}

As demonstrated above, a writing pulse causes the swap of Gaussian states of the signal photon to the mechanical oscillator and readout pulse at a later time results in the retrieval of the stored Gaussian states. Such states, namely coherent as well as squeezed states, can be efficiently transferred to the retrieved photon when the width of the writing and readout pulses is shorter than mechanical decoherence time. Again, to characterize the photon transfer through a levitated optomechanical system, we obtain the zero-delay second-order autocorrelation (see Appendix \ref{appendixC}) of the retrieved photon during readout pulse, $g^{2}(0)=\langle a_{re}^{\dagger^{2}}a_{re}^{2}\rangle\Big/\langle a^{\dagger}_{re}a_{re}\rangle^{2}$. As shown in Fig. \ref{fig7}(a), this function remains 1 [$g^{2}(0)\rightarrow 1$] as a function of the free evolution time, if the incident signal photon is in a coherent state.
However, for nonclassical squeezed states of the signal, the two-photon coincidence probability for the retrieved photon becomes less than 1 [$g^{2}(0)<1$], as depicted in Fig. \ref{fig7}(b). Thus, our protocol based on cavityless levitated optomechanics provides a versatile platform for the efficient transfer of single-photon Gaussian states. 
\subsection{Scattering matrix analysis}
\begin{figure}[ht!]
\begin{center}
\begin{tabular}{cc}
\includegraphics[scale=0.31]{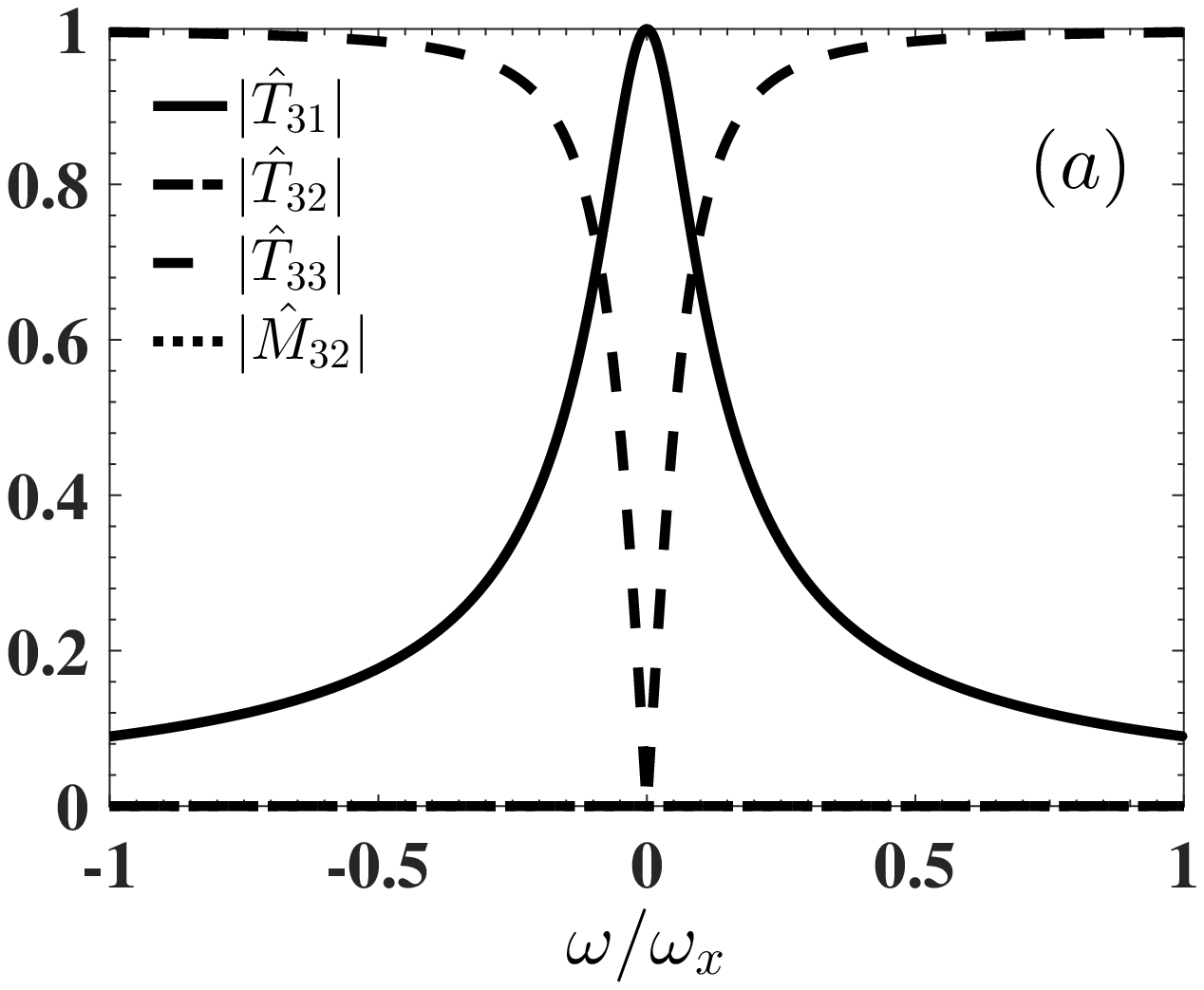} & \includegraphics[scale=0.31]{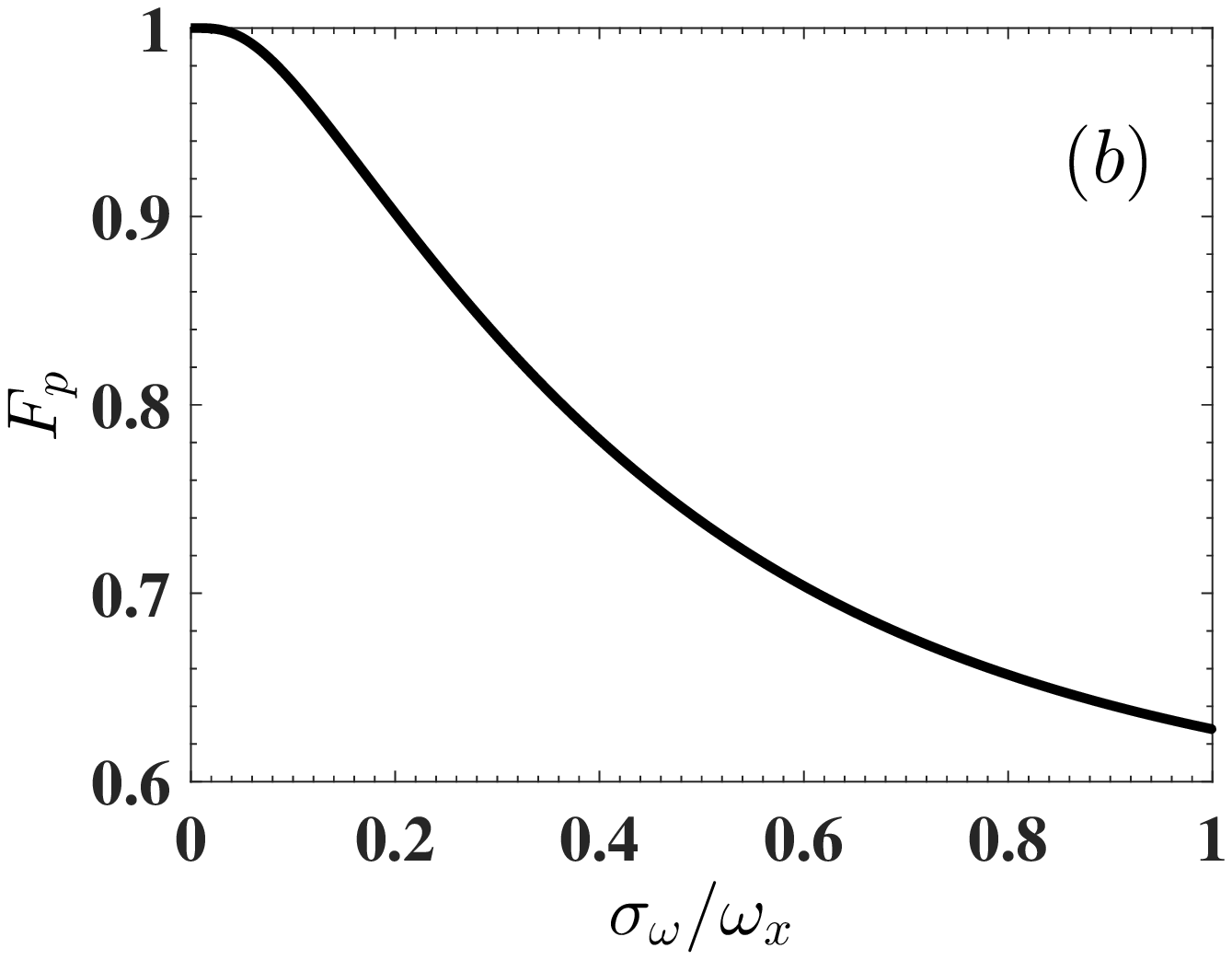}
\end{tabular}
\end{center}
\caption{(a) Transmission matrix coefficients vs frequency under impedance matching. (b) Pulse fidelity vs spectral width. All other parameters are same as in Fig. \ref{fig2}.}
\label{fig8}
\end{figure}

In the preceding analysis, we have described the transfer of Gaussian states associated with the signal. Here, we analyze the transmission of a single-photon signal pulse through the levitated optomechanical system.  Such traveling pulses can be transmitted between channels of quite different wavelengths \cite{kumar2018}. In the itinerant state transfer, an input state is centered around a single frequency. As a result of this, the transfer of signal pulse to the retrieved pulse can be viewed as a stationary scattering process and thus a high-fidelity transfer can be characterized by a set of requirements on scattering matrix \cite{safavi2011}. Now, to study itinerant state transfer, we consider a quantum input $a_{in}^{s}(t)$ for the signal $a_{s}$ and $b_{in}(t)$ and $a_{in}^{r}(t)$ are the noise operators with zero average. For constant effective couplings, the retrieved photon pulse at the output can be written as (see Appendix \ref{appendixD})  
\begin{align}
a_{re}(\omega)&=\hat{T}_{31}(\omega)a^{s}_{in}(\omega)+\hat{T}_{32}(\omega)b_{in,\mathcal{T}}(\omega)+\hat{T}_{33}(\omega)a^{r}_{in}(\omega)\nonumber\\
&+\hat{M}_{32}(\omega)b_{in,\mathcal{F}}(\omega)\label{eq9}\;,
\end{align}
where $\hat{T}_{31}(\omega)$ characterizes the transmission of input signal pulse $a^{s}_{in}(\omega)$ to output retrieved pulse $a_{re}(\omega)$, $\hat{T}_{32}(\omega)~(\hat{M}_{32}(\omega))$ represents the transmission of thermal (non-linear feedback) mechanical noise, and $\hat{T}_{33}(\omega)$ gives the contribution of the optical noise associated with readout pulse. For a high-fidelity transfer from  $a^{s}_{in}(\omega)$ to $a_{re}(\omega)$ over the bandwidth of the signal pulse, the transmission matrix coefficient $\hat{T}_{31}(\omega)\rightarrow 1$ as well as $\hat{T}_{32}(\omega),\hat{M}_{32}(\omega), \hat{T}_{33}(\omega)\rightarrow 0$. This is shown in Fig. \ref{fig8}(a) and it is clear that $|\hat{T}_{31}(\omega)|=1$, whereas other noise contributions are suppressed at $\omega=0$ in the transmission. The value of $\hat{T}_{31}$ at $\omega=0$ is $\hat{T}_{31}(0)=2\sqrt{C_{w}C_{r}}/(C_{w}+C_{r}+1)$, where $C_{w}=4G_{w}^{2}/\Gamma\mathcal{B}~(C_{r}=4G_{r}^{2}/\Gamma\mathcal{B}_{r})$ is the cooperativity associated with write (readout) process. It is to be noted $\hat{T}_{31}(0)$ attains a maximum value at an optimal transmission condition of impedance matching ($C_{w}=C_{r}$) \cite{dong2015,safavi2011}, thereby providing efficient transfer of a signal photon. 

Another important feature of the transmission matrix element is the transmission half-width $\Delta\omega$ defined as $|\hat{T}_{31}(\Delta\omega)|=|\hat{T}_{31}(0)|/2$. We find that, in the strong-coupling regime $G_{i}>\Gamma,\mathcal{B}$, the transmission half-width is determined by $4G_{i}^{2}/\mathcal{B}$. But in the weak coupling limit with $G_{i}<\Gamma,\mathcal{B}$, the transmission half-width is limited by the mechanical decay rate \cite{tian2012,tian2015}. This shows that for strong optomechanical coupling the input components $|\omega|\ll\Delta\omega$ can be efficiently transmitted since such frequency components of the signal photon remain immune to the mechanical decoherence. To characterize this, we define pulse fidelity \cite{tian2012} as  
\begin{align}
F_{p}&=\frac{\vert\int d\omega\langle a^{s}_{in}(\omega)\rangle\langle a_{re}(\omega)\rangle^{\ast}\vert^{2}}{\int d\omega|\langle a^{s}_{in}(\omega)\rangle|^{2}\int d\omega|\langle a_{re}(\omega)\rangle^{\ast}|^{2}}\;.\label{eq10}
\end{align}
Using Eq. (\ref{eq9}), $\langle a_{re}(\omega)\rangle=|\hat{T}_{31}(\omega)|\langle a^{s}_{in}(\omega)\rangle$ for frequency components and thus the pulse fidelity is determined by the properties of $\hat{T}_{31}(\omega)$. For illustration, we study the transmission of a Gaussian signal pulse $\langle a^{s}_{in}(\omega)\rangle=\frac{1}{\sqrt{\pi}\sigma_{\omega}}\exp\left(-\frac{\omega^{2}}{\sigma_{\omega}^{2}}\right)$, where $\sigma_{\omega}$ is the spectral width. If we use this pulse profile in Eq. (\ref{eq10}), then the pulse fidelity attains a high value for a narrow-bandwidth pulse, as shown in Fig. \ref{fig8}(b). Thus a quantum input signal pulse with spectral width $\sigma_{\omega}\ll\Delta\omega$ can be transmitted with high fidelity to the output without suffering from mechanical decoherence throughout the transmission. This facilitates an efficient photon-phonon-photon transfer in a levitated optomechanical system.
\section{Conclusion}
\label{Sec4}
We have studied the storage and retrieval of a single photon at a quantum level using levitated cavityless optomechanics. We have shown that, under experimental conditions, such a system is suitable for the efficient transfer of single-photon quantum states.  The effectiveness of the protocol was characterized in terms of the fidelity, the Wigner function, and the zero-delay second- order autocorrelation function. These quantities  were  explored to demonstrate a robust conversion of Gaussian states of the signal to the retrieved photon. We have found that our protocol remains relatively  immune to the mechanical decoherence in the presence of strong writing and readout pulses of duration smaller than mechanical decay. Further, a high fidelity photon-phonon-photon transfer was described in terms of the transmission of traveling photon pulses between channels of quite different frequencies. Our results indicate that levitated optomechanics may be useful for quantum networks.
\begin{acknowledgments}
The authors acknowledge support from the Office of Naval Research (N00014-17-1-2291).
\end{acknowledgments}   
\appendix
\section{CALCULATIONS FOR OPTOMECHANICAL STORAGE AND RETRIEVAL}
\label{appendixA}
To delineate a quantum memory for the storage and retrieval of single photon, we use Eqs. (\ref{eq4}) and \ref{eq5}) to obtain following equations for the second-order moments \cite{scully1997}:
\begin{align}
\frac{d}{dt}\langle a_{s}^{\dagger}a_{s}\rangle &=-2\mathcal{B}\langle a_{s}^{\dagger}a_{s}\rangle+iG_{i}\langle b^{\dagger}a_{s}\rangle-iG_{i}\langle a_{s}^{\dagger}b\rangle+4\mathcal{B}I_{in}\label{eqa1}\;,\\
\frac{d}{dt}\langle b^{\dagger}b\rangle &=-2\Gamma\langle b^{\dagger}b\rangle-iG_{i}\langle b^{\dagger}a_{s}\rangle+iG_{i}\langle a_{s}^{\dagger}b\rangle+2\gamma+2\mathcal{F}\label{eqa2}\;,\\
\frac{d}{dt}\langle b^{\dagger}a_{s}\rangle &=-\left[i\Delta+\mathcal{B}+\Gamma\right]\langle b^{\dagger}a_{s}\rangle+iG_{i}\langle a_{s}^{\dagger}a_{s}\rangle-iG_{i}\langle b^{\dagger}b\rangle\label{eqa3}\;,\\
\frac{d}{dt}\langle a_{s}^{\dagger}b\rangle &=-\left[-i\Delta+\mathcal{B}+\Gamma\right]\langle a_{s}^{\dagger}b\rangle-iG_{i}\langle a_{s}^{\dagger}a_{s}\rangle+iG_{i}\langle b^{\dagger}b\rangle\label{eqa4}\;.
\end{align}
To derive the above equations we have used the fact that $a_{s}\left(t\right),a_{s}^{\dagger}\left(t\right),b\left(t\right),b^{\dagger}\left(t\right)$ are not affected by noise $a^{s}_{in}\left(t^{\prime}\right),a_{in}^{s^{\dagger}}\left(t^{\prime}\right),b_{in}\left(t^{\prime}\right),b_{in}^{\dagger}\left(t^{\prime}\right)$ at different times \cite{orszag2008}. 
\section{CALCULATION OF COVARIANCE MATRIX}
\label{appendixB}
In order to calculate the covariance matrix associated with the retrieved photon optical quadratures, we write the following equation from Eqs. (\ref{eq4} and \ref{eq5}):-
\begin{align}
\dfrac{d}{dt}v(t)=\mathcal{C}v(t)+n(t)\;,\label{eqb1}
\end{align}
where
\begin{align}
v(t)=\begin{bmatrix}X\\Y\\Q\\P\end{bmatrix},n(t)=\begin{bmatrix}
X_{in}(t)\\Y_{in}(t)\\Q_{in}(t)\\P_{in}(t)
\end{bmatrix},\mathcal{C}=\begin{bmatrix}
-\mathcal{B} & 0 & 0 & G_{i}\\
0 & -\mathcal{B} & -G_{i} & 0\\
0 & G_{i} & -\Gamma & 0\\
-G_{i} & 0 & 0 & -\Gamma 
\end{bmatrix}\;.\label{eqb2}
\end{align}
Here, the quadratures of the optical and mechanical modes are given by $X=a_{s}^{\dagger}+a_{s}$, $Y=i\left(a_{s}^{\dagger}-a_{s}\right)$, 
$X_{in}=\sqrt{2\mathcal{B}}\left(\hat{a}_{in}^{s^{\dagger}}+\hat{a}_{in}^{s}\right)$, $Y_{in}=i\sqrt{2\mathcal{B}}\left(\hat{a}_{in}^{s^{\dagger}}-\hat{a}_{in}^{s}\right)$ and $Q=\hat{b}^{\dagger}+\hat{b}$, $P=i\left(\hat{b}^{\dagger}-\hat{b}\right)$, $Q_{in}=\sqrt{2\gamma}\left(\hat{b}_{in,\mathcal{T}}^{\dagger}+\hat{b}_{in,\mathcal{T}}\right)+\sqrt{2\mathcal{F}}\left(\hat{b}_{in,\mathcal{F}}^{\dagger}+\hat{b}_{in,\mathcal{F}}\right)$, $P_{in}=i\sqrt{2\gamma}\left(\hat{b}_{in,\mathcal{T}}^{\dagger}-\hat{b}_{in,\mathcal{T}}\right)+i\sqrt{2\mathcal{F}}\left(\hat{b}_{in,\mathcal{F}}^{\dagger}-\hat{b}_{in,\mathcal{F}}\right)$, respectively.

Now, the solution of Eq. (\ref{eqb1}) can be written as
\begin{align}
v(t)=M(t)v(0)+\int_{0}^{t}ds M(t-s)n(s)\;,\label{eqb3}
\end{align}
where $M(t)=\exp{\left(\mathcal{C} t\right)}$. Now, in the presence of writing pulse of duration $t_{1s}$, the above equation can be written in the simplified form as
\begin{widetext}
\begin{align}
X(t_{1s})&=\eta_{w}\Big[\cos\left(G_{w}t_{1s}\right)X(0)+\sin\left(G_{w}t_{1s}\right)P(0)\Big]+\eta_{w}\int_{0}^{t_{1s}}ds e^{\mathcal{B}s}\cos\left[G_{w}\left(t_{1s}-s\right)\right]X_{in}(s)\nonumber\\
&+\eta_{w}\int_{0}^{t_{1s}}dse^{\mathcal{B}s}\sin\left[G_{w}\left(t_{1s}-s\right)\right]P_{in}(s)\;,\label{eqb4}\\
Y(t_{1s})&=\eta_{w}\Big[\cos \left(G_{w}t_{1s}\right)Y(0)-\sin \left(G_{w}t_{1s}\right)Q(0)\Big]+\eta_{w}\int_{0}^{t_{1s}}ds e^{\mathcal{B}s}\cos \left[G_{w}\left(t_{1s}-s\right)\right]Y_{in}(s)\nonumber\\
&-\eta_{w}\int_{0}^{t_{1s}}dse^{\mathcal{B}s}\sin \left[G_{w}\left(t_{1s}-s\right)\right]Q_{in}(s)\;,\label{eqb5}\\
Q(t_{1s})&=e^{-\Gamma t_{1s}}\Big[\sin \left(G_{w}t_{1s}\right)Y(0)+\cos \left(G_{w}t_{1s}\right)Q(0)\Big]+e^{-\Gamma t_{1s}}\int_{0}^{t_{1s}}dse^{\Gamma s}\sin \left[G_{w}\left(t_{1s}-s\right)\right]Y_{in}(s)\nonumber\\
&+e^{-\Gamma t_{1s}}\int_{0}^{t_{1s}}dse^{\Gamma s}\cos \left[G_{w}\left(t_{1s}-s\right)\right]Q_{in}(s)\;,\label{eqb6}\\
P(t_{1s})&=-e^{-\Gamma t_{1s}}\Big[\sin \left(G_{w}t_{1s}\right)X(0)+e^{-\Gamma t_{1s}}\cos \left(G_{w}t_{1s}\right)P(0)\Big]-e^{-\Gamma t_{1s}}\int_{0}^{t_{1s}}dse^{\Gamma s}\sin \left[G_{w}\left(t_{1s}-s\right)\right]X_{in}(s)\nonumber\\
&+e^{-\Gamma t_{1s}}\int_{0}^{t_{1s}}dse^{\Gamma s}\cos \left[G_{w}\left(t_{1s}-s\right)\right]P_{in}(s)\;,\label{eqb7}
\end{align}
\end{widetext}
where $G_{w}$ is the effective optomechanical coupling due to writing pulse, $\mathcal{B}=\mathcal{B}_{s}+\mathcal{B}_{w}$ is the total optical damping due to write process including contributions from signal and writing pulse and $\eta_{w}=\exp{(-\mathcal{B}t_{1s})}$. After the writing process, the mechanical system is allowed to evolve freely for time $t_{f}$. During this stage, the mechanical environment intervenes in the process. Finally, during a readout pulse, the optical quadratures of the retrieved photon read as
\begin{widetext}
\begin{align}
X(t_{s})&=\eta_{r}\eta_{w}\Big[\cos\left(G_{r}t_{2s}\right)\cos\left(G_{w}t_{1s}\right)X(0)+\cos\left(G_{r}t_{2s}\right)\sin\left(G_{w}t_{1s}\right)P(0)\Big]\nonumber\\
&+\eta_{r}\eta_{f}\Big[-\sin\left(G_{r}t_{2s}\right)\sin \left(G_{w}t_{1s}\right)X(0)+\sin\left(G_{r}t_{2s}\right)\cos \left(G_{w}t_{1s}\right)P(0)\Big]\nonumber\\
&+\eta_{r}\eta_{w}\cos\left(G_{r}t_{2s}\right)\int_{0}^{t_{1s}}ds\left[e^{\mathcal{B}s}\cos\left[G_{w}\left(t_{1s}-s\right)\right]X_{in}(s)+e^{\mathcal{B}s}\sin\left[G_{w}\left(t_{1s}-s\right)\right]P_{in}(s)\right]\nonumber\\
&+\eta_{r}\eta_{f}\sin\left(G_{r}t_{2s}\right)\int_{0}^{t_{1s}}ds\left[-e^{\Gamma s}\sin \left[G_{w}\left(t_{1s}-s\right)\right]X_{in}(s)+e^{\Gamma s}\cos \left[G_{w}\left(t_{1s}-s\right)\right]P_{in}(s)\right]\nonumber\\
&+\eta_{r}\int_{0}^{t_{2s}}ds\left[e^{\mathcal{B}_{r}s}\cos\left[G_{r}\left(t_{2s}-s\right)\right]X^{r}_{in}(s)+e^{\mathcal{B}_{r}s}\sin\left[G_{r}\left(t_{2s}-s\right)\right]P_{in}(s)\right]\;,\label{eqb8}\\
Y(t_{s})&=\eta_{r}\eta_{w}\Big[\cos\left(G_{r}t_{2s}\right)\cos\left(G_{w}t_{1s}\right)Y(0)-\cos\left(G_{r}t_{2s}\right)\sin\left(G_{w}t_{1s}\right)Q(0)\Big]\nonumber\\
&-\eta_{r}\eta_{f}\Big[\sin\left(G_{r}t_{2s}\right)\sin \left(G_{w}t_{1s}\right)Y(0)+\sin\left(G_{r}t_{2s}\right)\cos \left(G_{w}t_{1s}\right)Q(0)\Big]\nonumber\\
&+\eta_{r}\eta_{w}\cos\left(G_{r}t_{2s}\right)\int_{0}^{t_{1s}}ds\left[e^{\mathcal{B}s}\cos \left[G_{w}\left(t_{1s}-s\right)\right]Y_{in}(s)-e^{\mathcal{B}s}\sin \left[G_{w}\left(t_{1s}-s\right)\right]Q_{in}(s)\right]\nonumber\\
&-\eta_{r}\eta_{f}\sin\left(G_{r}t_{2s}\right)\int_{0}^{t_{1s}}ds\left[e^{\Gamma s}\sin \left[G_{w}\left(t_{1s}-s\right)\right]Y_{in}(s)+e^{\Gamma s}\cos \left[G_{w}\left(t_{1s}-s\right)\right]Q_{in}(s)\right]\nonumber\\
&+\eta_{r}\int_{0}^{t_{2s}}ds\left[e^{\mathcal{B}_{r}s}\cos \left[G_{w}\left(t_{2s}-s\right)\right]Y^{r}_{in}(s)-e^{\mathcal{B}_{r}s}\sin \left[G_{w}\left(t_{2s}-s\right)\right]Q_{in}(s)\right]\;,\label{eqb9}
\end{align}
\end{widetext}
where $\eta_{f}=\exp\left[-\Gamma\left(t_{1s}+t_{f}\right)\right]$, $\eta_{r}=\exp\left(-\mathcal{B}_{r}t_{2s}\right)$, $t_{2s}$ is the duration of the readout pulse, $t_{s}$ is the total transfer time, $G_{r}$ is the effective optomechanical during readout pulse, and $\mathcal{B}_{r}$ is the optical damping during readout process. Now, the fidelity can be  calculated by using the following covariance matrix
\begin{align}
V=\begin{bmatrix}
V_{X_{t_{s}}X_{t_{s}}} & V_{X_{t_{s}}Y_{t_{s}}}\\
V_{Y_{t_{s}}X_{t_{s}}} & V_{Y_{t_{s}}Y_{t_{s}}}
\end{bmatrix}\;,\label{eqb10}
\end{align}
where the elements of the covariance matrix are defined as $V_{\xi_{i}\xi_{j}}=\frac{1}{2}\langle\xi_{i}\xi_{j}+\xi_{j}\xi_{i}\rangle-\langle\xi_{i}\rangle\langle\xi_{j}\rangle$ and these can further be written by using Eqs. (\ref{eqb8}) and \ref{eqb9}). In order to derive covariance matrix elements, we require $\langle X(0)X(0)\rangle=(\alpha+\alpha^{\ast})^{2}+e^{-2r}$, and  $\langle P(0)P(0)\rangle=(2\langle N\rangle+1)$. It turns  out that $V_{Y(t_{s})Y(t_{s})}$ can be expressed in terms of $V_{X(t_{s})X(t_{s})}$ just by replacing $r$ to $-r$ ; also, $V_{X(t_{s})Y(t_{s})}=V_{Y(t_{s})X(t_{s})}=0$.
\section{CALCULATION OF ZERO-DELAY SECOND-ORDER AUTOCORRELATION FUNCTION ($g^{2}(0)$)}
\label{appendixC}
Let us again start from Eqs. (\ref{eq4}) and \ref{eq5}). The solution of these equations can be written as
\begin{align}
f(t)=e^{\mathcal{M} t}f(0)+\int_{0}^{t}ds~e^{\left[\mathcal{M}(t-s)\right]}f_{in}(s)\;,\label{eqc1}
\end{align}
where
\begin{align}
f(t)=\begin{bmatrix}
a_{s}\\b
\end{bmatrix},\mathcal{M} = \begin{bmatrix}
-\mathcal{B} & -i G_{w}\\
-i G_{w} & -\Gamma
\end{bmatrix},f_{in}=\begin{bmatrix}
a^{s}_{in}\\b_{in,\mathcal{T}}+b_{in,\mathcal{F}}
\end{bmatrix}\;.\label{eqc2}
\end{align}
Now in the photon-phonon-photon transfer, the states of the signal are transferred to the mechanical mode by employing a writing pulse of duration $t_{1s}$. Then the system is evolved freely for time $t_{f}$. Finally, the stored photon is retrieved at a later time by using a readout pulse of span $t_{2s}$ and can be written as  
\begin{align}
a_{re}(t_{s})&=\eta_{r}\cos(G_{r}t_{2s})a_{r}(0)\nonumber\\
&-\eta_{r}\eta_{f}\sin(G_{r}t_{2s})\sin(G_{w}t_{1s})a_{s}(0)\nonumber\\
&-i\eta_{r}\eta_{f}\sin(G_{r}t_{2s})\cos(G_{w}t_{1s})b(0)\nonumber\\
&-i\eta_{r}\eta_{f}\sin(G_{r}t_{2s})\int\limits_{0}^{t_{1s}}ds e^{\Gamma s}\cos\left[G_{w}(t_{1s}-s)\right]b_{in,\mathcal{T}}^{s}(s)\nonumber\\
&-i\eta_{r}\eta_{f}\sin(G_{r}t_{2s})\int\limits_{0}^{t_{1s}}ds e^{\Gamma s}\cos\left[G_{w}(t_{1s}-s)\right]b_{in,\mathcal{F}}^{s}(s)\nonumber\\
&-\eta_{r}\eta_{f}\sin(G_{r}t_{2s})\int\limits_{0}^{t_{1s}}ds e^{\Gamma s}\sin\left[G_{w}(t_{1s}-s)\right]a_{in}^{s}(s)\nonumber\\
&+\eta_{r}\int\limits_{0}^{t_{2s}}ds e^{\mathcal{B}_{r}s}\cos\left[G_{r}(t_{2s}-s)\right]a_{in}^{r}(s)\;,\label{eqc3}
\end{align}
where $\eta_{f}$ and $\eta_{r}$ are defined in Appendix \ref{appendixB}. Now, from Eq. (\ref{eqc3}), we can write $\langle a_{re}^{\dagger}a_{re}\rangle$ and it can further be simplified in the following by solving the integration
\begin{align}
\langle a_{re}^{\dagger}a_{re}\rangle &=\eta_{r}^{2}\eta_{f}^{2}\sin^{2}\left(G_{r}t_{2s}\right)\sin^{2}\left(G_{w}t_{1s}\right)\left(|\alpha|^{2}+\sinh^{2}(r)\right)\nonumber\\
&+\eta_{r}^{2}\eta_{f}^{2}\sin^{2}\left(G_{r}t_{2s}\right)\cos^{2}\left(G_{w}t_{1s}\right)\langle N\rangle+2\mathcal{B}_{r}\eta_{2}^{2}\mathcal{G}_{33}\nonumber\\
&+2\eta_{r}^{2}\eta_{f}^{2}\sin^{2}\left(G_{r}t_{2s}\right)\Big[\mathcal{B}\mathcal{G}_{11}
+\left(\gamma+\mathcal{F}\right)\mathcal{G}_{22}\Big]\;,\label{eqc4}
\end{align}
where, we have used that for a squeezed coherent state of the signal $\langle a_{s}^{\dagger}(0)a_{s}(0)\rangle=|\alpha|^{2}+\sinh^{2}(r)$, for thermal state of the mechanical system $\langle b^{\dagger}(0)b(0)\rangle =\langle N\rangle$ and for the vacuum state of the readout $\langle a_{r}^{\dagger}(0)a_{r}(0)\rangle =0$. Also $\mathcal{G}_{11},\mathcal{G}_{22}$, in Eq. (\ref{eqc4}) can be written as,   
\begin{align}
\mathcal{G}_{11}&=-\frac{G_{w}^{2}\mathcal{N}_{1}+\Gamma^{2}\mathcal{N}_{2}+G_{w}\Gamma\mathcal{N}_{3}}{\mathcal{R}}\;,\label{eqc5}\\
\mathcal{G}_{22}&=-\frac{G_{w}^{2}\mathcal{N}_{1}+\Gamma^{2}\left(2\mathcal{N}_{1}-\mathcal{N}_{2}\right)-G_{w}\Gamma\mathcal{N}_{3}}{\mathcal{R}}\;,\label{eqc6}
\end{align}
where $\mathcal{N}_{1}=\left(1-e^{2\Gamma t_{1s}}\right)$, $\mathcal{N}_{2}=1-\cos\left(2G_{w}t_{1s}\right)$, $\mathcal{N}_{3}=\sin\left(2G_{w}t_{1s}\right)$ and $\mathcal{R}=4\Gamma\left(G_{w}^{2}+\Gamma^{2}\right)$. Further, $\mathcal{G}_{33}$ can be written from $\mathcal{G}_{22}$ in Eq. (\ref{eqc6}) just by replacing $\Gamma,G_{w},t_{1s}$ by $\mathcal{B}_{r},G_{r},t_{2s}$, respectively. 

Similarly, from Eq. (\ref{eqc3}) we can write $\langle a_{re}^{\dagger^{2}}a_{re}^{2}\rangle$. This factor can be simplified by using a moment-factoring theorem \cite{razavi2009,shapiro1994} and is given by
\begin{align}
\langle a_{re}^{\dagger^{2}}a_{re}^{2}\rangle&=\eta_{r}^{4}\eta_{f}^{4}\sin^{4}(G_{r}t_{2s})\sin^{4}(G_{w}t_{1s})\mathcal{U}+8\mathcal{B}_{r}^{2}\eta_{r}^{4}\mathcal{G}_{33}^{2}\nonumber\\
&+\eta_{r}^{4}\eta_{f}^{4}\sin^{4}(G_{r}t_{2s})\cos^{4}(G_{w}t_{1s})\langle N\rangle^{2}\nonumber\\
&+8\left(\gamma^{2}+\mathcal{F}^{2}\right)\eta_{r}^{4}\eta_{f}^{4}\sin^{4}(G_{r}t_{2s})\mathcal{G}_{22}^{2}\nonumber\\
&+8\mathcal{B}^{2}\eta_{r}^{4}\eta_{f}^{4}\sin^{4}(G_{r}t_{2s})\mathcal{G}_{11}^{2}\;,\label{eqc7}
\end{align}
where, $\mathcal{U}=\sinh^{2}(r)\cosh^{2}(r)-(\alpha^{2}+\alpha^{\ast^{2}})\sinh(r)\cosh(r)+2\sinh^{4}(r)+4|\alpha|^{2}\sinh^{2}(r)+|\alpha|^{4}$. Finally, using Eqs. (\ref{eqc4}) and \ref{eqc7}), we can write $g^{2}(0)$.
\section{CALCULATIONS OF SCATTERING MATRIX}
\label{appendixD}
In this analysis, we describe how the traveling photon pulses can be transmitted from input and output channels of distinctly different wavelengths.  To do so, let us consider a quantum input $a_{in}^{s}(t)$ for the signal $a_{s}$ and $b_{in}(t)$ and $a_{in}^{r}(t)$ are the noise operators with zero average. Thus the quantum Langevin equations describing the present system can be written as
\begin{align}
\frac{du(t)}{dt}&=Au(t)+Ku_{in}(t)+\mathcal{S}_{\mathcal{F}}\mathcal{B}_{in,\mathcal{F}}\;,\label{eqd1}
\end{align}
where
\begin{align}
u(t)&=\begin{bmatrix}
a_{s}(t)\\b(t)\\a_{r}(t)
\end{bmatrix},A=\begin{bmatrix}
A_{11} &-iG_{w}&0\\
-iG_{w}&-\Gamma&-iG_{r}\\
0&-iG_{r}&A_{33}
\end{bmatrix}\;,\label{eqd2}\\
u_{in}(t)&=\begin{bmatrix}
a_{in}^{s}(t)\\b_{in,\mathcal{T}}(t)\\a_{in}^{r}(t)
\end{bmatrix},K=\begin{bmatrix}
\sqrt{2\mathcal{B}}&0&0\\
0&\sqrt{2\gamma}&0\\
0&0&\sqrt{2\mathcal{B}_{r}}
\end{bmatrix}\;,\label{eqd3}\\
\mathcal{B}_{in,\mathcal{F}}(t)&=\begin{bmatrix}
0\\b_{in,\mathcal{F}}(t)\\0
\end{bmatrix},\mathcal{S}_{\mathcal{F}}=\begin{bmatrix}
0&0&0\\
0&\sqrt{2\mathcal{F}}&0\\
0&0&0
\end{bmatrix}\;.\label{eqd4}
\end{align}
where detuning is defined by $\Delta_{i}=\omega_{s}-\omega_{i}~(i=,w,r)$, $A_{11} = -(i\Delta_{1}+\mathcal{B})$, and $A_{33} = -(i\Delta_{2}+\mathcal{B}_{r})$.

Now, taking Fourier Transform of Eq. (\ref{eqd1}), we get
\begin{align}
u(\omega)=\Big(i\omega I-A\Big)^{-1}Ku_{in}(\omega)+\Big(i\omega I-A\Big)^{-1}\mathcal{S}_{\mathcal{F}}\mathcal{B}_{in,\mathcal{F}}\;.\label{eqd5}
\end{align}
Further, using $u_{out}(\omega)=Ku(\omega)-u_{in}(\omega)$, we get
\begin{align}
u_{out}(\omega)=\hat{T}(\omega)u_{in}(\omega)+\hat{M}(\omega)\mathcal{B}_{in,\mathcal{F}}(\omega)\;,\label{eqd6}
\end{align}
where $\hat{T}(\omega)=K\Big(i\omega I-A\Big)^{-1}K-I$ and $\hat{M}(\omega)=K\Big(i\omega I-A\Big)^{-1}\mathcal{S}_{\mathcal{F}}$. The retrieved pulse ($a^{r}_{out}(\omega)$) can be written from Eq. (\ref{eqd1}) and depends on transmission matrix elements $\hat{T}_{31}(\omega),\hat{T}_{32}(\omega),\hat{T}_{33}(\omega),\hat{M}_{32}(\omega)$ as is given in Eq. (\ref{eq9}). These transmission matrix elements can be written as
\begin{align}
\hat{T}_{31}(\omega)&=-\frac{\sqrt{2\mathcal{B}}\sqrt{2\mathcal{B}_{r}}G_{w}G_{r}}{\mathcal{I}}\;,\label{eqd7}\\
\hat{T}_{33}(\omega)&=\frac{2\mathcal{B}_{r}\Big[(i\omega+\Gamma)\left[i(\omega+\Delta_{1})+\mathcal{B}\right]+G_{w}^{2}\Big]}{\mathcal{I}}-1\label{eqd8}\\
\hat{T}_{32}(\omega)&=-\frac{i\sqrt{2\mathcal{B}_{r}}\sqrt{2\gamma}G_{r}\left[i(\omega+\Delta_{1})+\mathcal{B}\right]}{\mathcal{I}}\label{eqd9}\\
\hat{M}_{32}(\omega)&=-\frac{i\sqrt{2\mathcal{B}_{r}}\sqrt{2\mathcal{F}}G_{r}\left[i(\omega+\Delta_{1})+\mathcal{B}\right]}{\mathcal{I}}\label{eqd10}
\end{align}
where $\mathcal{I}=\Big[i(\omega+\Delta_{1})+\mathcal{B}\Big]\Big[(i\omega+\Gamma)\left[i(\omega+\Delta_{2})+\mathcal{B}_{r}\right]+G_{r}^{2}\Big]+G_{w}^{2}\left[i(\omega+\Delta_{2})+\mathcal{B}_{r}\right]$. 

To transmit an input signal pulse $a^{s}_{in}(\omega)$ to the output retrieved pulse $a_{re}(\omega)$, two conditions need to hold. (1) The information from the input channel needs to be efficiently transferred to the output channel. This condition demands that $|\hat{T}_{31}(\omega)|\rightarrow 1$.  (2) The noise needs to be blocked from entering in this process. This condition requires that $|\hat{T}_{32}(\omega)|, |\hat{T}_{33}(\omega)|, |\hat{M}_{32}(\omega)|\rightarrow 0$. These conditions are obeyed at $\Delta_{i}=-\omega_{m}$ as $\omega\rightarrow 0$. Further,  $|\hat{T}_{31}(\omega)|\rightarrow 1$ as $\omega\rightarrow 0$ under optimal transmission condition $\mathcal{B}_{r}G_{w}^{2}=\mathcal{B}G_{r}^{2}$,  i.e., when impedance matching $C_{w}=C_{r}$ is obeyed. Here, $C_{w}=\frac{4G_{w}^{2}}{\mathcal{B}\Gamma}~\left(C_{r}=\frac{4G_{r}^{2}}{\mathcal{B}_{r}\Gamma}\right)$ is the cooperativity associated with the write (readout) process. Moreover, under this condition $|\hat{T}_{32}(\omega)|, |\hat{T}_{33}(\omega)|, |\hat{M}_{32}(\omega)|\rightarrow 0$ as $\omega\rightarrow 0$.

Further, the pulse fidelity equivalent to Eq. (\ref{eq10}) can be defined as
\begin{align}
F_{p}&=\frac{\vert\int dt\langle a^{s}_{in}(t)\rangle\langle a_{re}(t)\rangle^{\ast}\vert^{2}}{\int dt|\langle a^{s}_{in}(t)\rangle|^{2}\int dt|\langle a_{re}(t)\rangle^{\ast}|^{2}}\;,\label{eqd11}
\end{align}
where, the output retrieved pulse $a_{re}(t)$ can be calculated by integrating over frequency components $a_{re}(t)=\int d\omega\langle a_{re}(\omega)\rangle e^{i\omega t}$.

\bibliographystyle{elsarticl-num}

\begin{thebibliography}{50}
\bibitem{aspelmeyer2014} M. Aspelmeyer, T. J. Kippenberg, and F. Marquardt, Cavity optomechanics, Rev. Mod. Phys. \textbf{86}, 1391 (2014).
\bibitem{meystre2013} P. Meystre, A short walk through quantum optomechanics, Ann. Phys. (Berlin) \textbf{525}, 215 (2012).
\bibitem{cdong2015} C. Dong, Y. Wang, and H. Wang, Optomechanical interfaces for hybrid quantum networks, Nat. Phys. Rev. \textbf{2}, 510 (2015).
\bibitem{felicetti2017} S. Felicetti, S. Fedortchenko, R. Rossi Jr., S. Ducci, I. Favero, T. Coudreau, and P. Milman, Quantum communication between remote mechanical resonators, Phys. Rev. A \textbf{95}, 022322 (2017).
\bibitem{fiore2011} V. Fiore, Y. Yang, M. C.  Kuzyk, R. Barbour, L. Tian, and H. Wang, Storing optical information as a
mechanical excitation in a silica optomechanical resonator, Phys. Rev. Lett. \textbf{107}, 133601 (2011).
\bibitem{fiore2013} V. Fiore, C.  Dong, M. C. Kuzyk, and H. Wang, Optomechanical light storage in a silica microresonator,
Phys. Rev. A \textbf{87}, 023812 (2013).
\bibitem{tian2010} L. Tian,and H. Wang, Optical wavelength conversion of quantum states with optomechanics, Phys. Rev. A \textbf{82}, 053806 (2010).
\bibitem{dong2015} C. Dong, V.  Fiore, M. C. Kuzyk, L. Tian, and H. Wang, Optical wavelength conversion via optomechanical
coupling in a silica resonator, Ann. der Phys. \textbf{527}, 100 (2015).
\bibitem{tian2015} L. Tian, Optoelectromechanical transducer: Reversible conversion between microwave and optical photons, Ann. Phys. (Berlin) \textbf{527}, 1 (2015).
\bibitem{stannigel2010} K. Stannigel, P. Rabl, A.S. Sorensen, P. Zoller, and M. D. Lukin, Optomechanical transducers for
long-distance quantum communication, Phys. Rev. Lett. \textbf{105}, 220501 (2010).
\bibitem{safavi2011} A. H. Safavi-Naeini, and O. Painter, Proposal for an optomechanical traveling wave phononphoton translator,
New J. Phys. \textbf{13}, 013017 (2011).
\bibitem{tian2012} L. Tian, Adiabatic state conversion and pulse transmission in optomechanical systems, Phys. Rev. Lett. \textbf{108}, 153604 (2012).
\bibitem{wang2012} Y. D. Wang, and A. A. Clerk, Using interference for high fidelity quantum state transfer in optomechanics, Phys. Rev. Lett. \textbf{108}, 153603 (2012).
\bibitem{clerk2012} Y. D. Wang, and A. A. Clerk, Using dark modes for high-fidelity optomechanical
quantum state transfer, New. J. Phys. \textbf{14}, 105010 (2012).
\bibitem{palomaki2013} T. A. Palomaki, J. W. Harlow, J. D. Teufel, R. W. Simmonds, and K. W. Lehnert, Coherent state transfer between itinerant microwave fields and a  mechanical oscillator, Nat. Lett. \textbf{495}, 210 (2013).
\bibitem{galland2014} C. Galland, N. Sangouard, N. Piro, N. Gisin, and T. J. Kippenberg, Heralded single-phonon preparation, storage, and readout in cavity optomechanics, Phys. Rev. Lett. \textbf{112}, 143602 (2014).
\bibitem{filip2015} R. Filip, and A. A. Rakhubovsky, Transfer of non-Gaussian quantum states of mechanical oscillator of light, Phys. Rev. A. \textbf{92}, 053804 (2015).
\bibitem{filip2017} A. A. Rakhubovsky, and R. Filip, Photon-phonon-photon transfer in optomechanics, Sci. Rep. \textbf{7}, 46764 (2017).
\bibitem{caprara2016} V. C. Vivoli, T. Barnea, C. Galland, and N. Sangouard, Proposal for an optomechanical bell test, Phys. Rev. Lett. \textbf{116}, 070405 (2016).
\bibitem{anderson2018} M. D. Anderson, S. T. Velez, K. Seibold, H. Flayac, V. Savona, N. Sangouard, and C. Galland, Two-color pump-probe measurement of photonic quantum correlations mediated by a single phonon, Phys. Rev. Lett. \textbf{120}, 233601 (2018).
\bibitem{connell2010} A. D. O'Connell, M. Hofheinz, M.  Ansmann, R. C.  Bialczak, M.  Lenander, E.  Lucero, M.  Neeley, D. Sank,
H. Wang, M. Weides, J. Wenner, J. M.  Martinis, and A. N. Cleland,  Quantum ground state and
single-phonon control of a mechanical resonator, Nature \textbf{464}, 697 (2010).
\bibitem{chan2011} J. Chan, T. P. M. Alegre, A. H. Safavi-Naeini, J. T.  Hill, A. Krause, S. Gr\"{o}blacher, M. Aspelmeyer,
and O. Painter, Laser cooling of a nanomechanical oscillator into its quantum ground state, Nature \textbf{478},
89 (2011).
\bibitem{anetsberger2010} G. Anetsberger, E. Gavartin, O. Arcizet, Q. P. Unterreithmeier, E. M. Weig, M. L. Gorodetsky, J. P. Kotthaus, and T. J. Kippenberg, Measuring nanomechanical motion with an imprecision below the standard quantum limit, Phys. Rev. A \textbf{82}, 061804(R) (2010).
\bibitem{groblacher2009} S. Gr\"{o}blacher, K. Hammerer, M. R. Vanner, and M.  Aspelmeyer, Observation of strong coupling between
a micromechanical resonator and an optical cavity field, Nature \textbf{460}, 724 (2009).
\bibitem{tpalomaki2013} T. A. Palomaki, J. D. Teufel, R. W. Simmonds, and K. W. Lehnert, Entangling mechanical motion with microwave fields, Science \textbf{342}, 710 (2013). 
\bibitem{pirkkalainen2015} J.-M. Pirkkalainen, E. Damsk\"{a}gg, M. Brandt, F. Massel, and M. A. Sillanp\"{a}\"{a}, Squeezing of quantum noise of motion in a micromechanical resonator, Phys. Rev. Lett. \textbf{115}, 243601 (2015).
\bibitem{agarwal2010} G. S. Agarwal, and S.  Huang, Electromagnetically induced transparency in mechanical effects of light,
Phys. Rev. A \textbf{81}, 041803 (2010).
\bibitem{weis2010} S. Weis,  R. Rivi\`{e}re, S. Del\'{e}glise, E. Gavartin, O.  Arcizet, A.  Schliesser, and T. J. Kippenberg, Optomechanically induced transparency, Science \textbf{330}, 1520 (2010).
\bibitem{zhou2013} X. Zhou, F. Hocke, A. Schliesser, A. Marx, H. Huebl, R. Gross, and  T. J. Kippenberg, Slowing, advancing and switching of microwave signals using circuit nanoelectromechanics, Nat. Phys. \textbf{9}, 179 (2013).
\bibitem{neukirch2015} L. P. Neukirch, and A. N. Vamivakas, Nano-optomechanics with optically levitated nanoparticles, Contemp. Phys. \textbf{56}, 48 (2015).
\bibitem{chang2009} D. E. Chang, C. A. Regal, S. B. Papp, D. J. Wilson, J. Ye, O. Painter, H. J. Kimble, and P. Zoller, Cavity optomechanics using optically levitated nanosphere, Proc. Natl. Acad. Sci. U.S.A. \textbf{107}, 1005 (2009).
\bibitem{yin2013} Z. Q. Yin, A. A. Geraci, and T. Li, Optomechanics of levitated dielectric particles, Int. J. Mod. Phys. B \textbf{27}, 1330018 (2013). 
\bibitem{romero2012} O. Romero-Isart, L. Clemente, C. Navau, A. Sanchez, and J. I. Cirac, Quantum magnetomechanics with levitating superconducting microspheres, Phys. Rev. Lett. \textbf{109}, 147205 (2012).
\bibitem{cirio2012} M. Cirio, G. K. Brennen, and J. Twamley, Quantum magnetomechanics: Ultrahigh-Q-levitated mechanical
oscillators, Phys. Rev. Lett. \textbf{109}, 147206 (2012).
\bibitem{li2011} T. Li, S. Kheifets, and M. G. Raizen, Millikelvin cooling of an optically trapped microsphere in vacuum, Nat.
Phys. \textbf{7}, 527 (2011).
\bibitem{gieseler2012} J. Gieseler, B. Deutsch, R. Quidant, and L. Novotny, Subkelvin parametric feedback cooling of a laser-trapped nanoparticle, Phys. Rev. Lett. \textbf{109}, 103603 (2012).
\bibitem{arita2013} Y. Arita, M. Mazilu, and K. Dholakia, Laser-induced rotation and cooling of a trapped microgyroscope in vacuum, Nat. Commun. \textbf{4}, 2374 (2013).
\bibitem{jain2016} V. Jain, J. Gieseler, C. Moritz, C. Dellago, R. Quidant, and L. Novotny, Direct measurement of photon recoil from a levitated nanoparticle, Phys. Rev. Lett. \textbf{116}, 243601 (2016).
\bibitem{frimmer2016} M. Frimmer, J. Gieseler, and L. Novotny, Cooling mechanical oscillators by coherent control, Phys. Rev. Lett. \textbf{117}, 163601 (2016). 
\bibitem{rodenburg2016} B. Rodenburg, L. P. Neukirch, A. N. Vamivakas, and M. Bhattacharya, Quantum model of cooling and force
sensing with an optically trapped nanoparticle, Optica \textbf{3}, 318 (2016).
\bibitem{moore2014} D. C. Moore, A. D. Rider, and G. Gratta, Search for millicharged particles using optically levitated microspheres, Phys. Rev. Lett. \textbf{113}, 251801 (2014).
\bibitem{ranjit2015} G. Ranjit, D. P. Atherton, J. H. Stutz, M. Cunningham, and A. A. Geraci, Attonewton force detection using
microspheres in a dual-beam optical trap in high vacuum, Phys. Rev. A \textbf{91}, 051805(R) (2015).
\bibitem{frimmer2017} M. Frimmer, K. Luszcz, S. Ferreiro, V. Jain, E. Hebestreit, and L. Novotny, Controlling the net charge on a nanoparticle optically levitated in vacuum, Phys. Rev. A \textbf{95}, 061801(R) (2017).
\bibitem{kumar2017} P. Kumar, and M. Bhattacharya, Magnetometry via spin-mechanical coupling in levitated optomechanics, Opt. Exp. \textbf{25}, 19568 (2017).
\bibitem{hebestreit2018} E. Hebestreit, R. Reimann, M. Frimmer, and L. Novotny, Measuring the internal temperature of a levitated nanoparticle in high vacuum, Phys. Rev. A \textbf{97}, 043803 (2018).
\bibitem{monteiro2018} F. Monteiro, S. Ghosh, E. C. van Assendelft, and D. C. Moore, Optical rotation of levitated spheres in high vacuum, Phys. Rev. A \textbf{97}, 051802(R) (2018).
\bibitem{isart2010} O. Romero-Isart, M. L. Luan, R. Quidant, and J. I. Cirac, Toward quantum superposition of living organisms,
New J. Phys. \textbf{12}, 033015 (2010).
\bibitem{isart2011} O. Romero-Isart, A. C. Pflanzer, F. Blaser, R. Kaltenbaek, N. Kiesel, M. Aspelmeyer, and J. I. Cirac, Large quantum superpositions and interference of massive nanometer-sized objects, Phys. Rev. Lett. \textbf{107}, 020405 (2011).
\bibitem{kumar2018} P. Kumar, and M. Bhattacharya, Storage and retrieval of optical information in levitated cavityless optomechanics, Proc. SPIE, Quant, Info. Sci., Sens., and Compu. X \textbf{10660}, 106600K (2018).
\bibitem{lpneukirch2015} L. P. Neukirch, E. Hartmaan, J. M. Rosenholm, and A. N. Vamivakas, Multi-dimensional single-spin nanooptomechanics with levitated nanodiamond, Nat. Phton. \textbf{9}, 653 (2015).
\bibitem{isar2008} A. Isar, Quantum fidelity for Gaussian states describing the evolution of open systems, Eur. Phys. J. Special Topics \textbf{160}, 225 (2008).
\bibitem{scully1997} M. O. Scully and M. S. Zubairy, \textit{Quantum Optics} (Cambridge University, Cambridge, 1997).
\bibitem{ge2016} W. Ge, B. Rodenburg, and M. Bhattacharya, Feedback-induced bistability of an optically levitated nanoparticle: A
Fokker-Planck treatment, Phys. Rev. A \textbf{94}, 023808 (2016).
\bibitem{pflanzer2012} A. C. Pflanzer, O. R. Isart, and J. I. Cirac, Master-equation approach to optomechanics with arbitrary dielectrics, Phys. Rev. A \textbf{86}, 013802 (2012).
\bibitem{breuer2002} H. -P. Breuer, and F. Pertuccione, \textit{The Theory of Open Quantum Systems} (Oxford University Press, 2002).
\bibitem{walls1970} D. F. Walls, Higher order effects in the master equation for coupled systems, Z. Physik \textbf{234}, 231 (1970).
\bibitem{schwendimann1972} P. Schwendimann, Interference of coherent and incoherent interactions in the master equation for open systems, Z. Physik \textbf{251}, 244 (1972).
\bibitem{linearization_reason} The linearized Hamiltonian in Eq. (\ref{eq3}) can be derived from Eq. (\ref{eq2}) by using mean-field approximation in which the intense control field can be treated classically and the linearization of the optomechanical interaction is performed with respect to the signal field i.e. we can write $a = \sqrt{n_{i}}e^{-i\omega_{i}t}+a_{s}~(i = w,r)$. During readout process, a same type of linearization can also be performed with respect to readout pulse.
\bibitem{damping_reason} During writing process the mechanical damping becomes $\Gamma_{w} = \gamma_{g}+\gamma_{w}+\gamma_{s}+\gamma_{t}+\delta\Gamma$, where $\gamma_{j}=\frac{5\mathcal{A}_{j}\omega_{x}}{\omega_{j}}~(j=t,w,s,r)$ is the rate associated with the radiation damping \cite{jain2016}. Here, $\mathcal{A}_{j}~(j=w,s,r)$ is the heating rate of the nanoparticle due to scattering of optical field and is obtained after the linearization of the second contribution in the optical scattering term written in Eq. (\ref{eq1}). Now, during readout process it takes the form of $\Gamma_{r} = \gamma_{g}+\gamma_{r}+\gamma_{t}+\delta\Gamma$. However, the main contribution to mechanical damping comes from nonlinear feedback and gas damping. In this regard, we can write $\Gamma_{w}\approx\Gamma_{r}\approx\gamma_{g} + \delta\Gamma = \Gamma$. 
\bibitem{qu2012} K. Qu, and G. S. Agarwal, Optical memories and transduction of fields in double cavity optomechanical systems, arXiv: 1210.4067 (2012).
\bibitem{orszag2008} M. Orszag, \textit{Quantum Optics} (Springer-Verlag, Berlin, 2008).
\bibitem{razavi2009} M. Razavi, I. S\"{o}llner, E. Bocquillon, C. Couteau, R. Laflamme, and G. Weihs, Characterizing heralded single-photon sources with imperfect measurement devices, Journal of Physics B: At., Mol. and Opt. Phys. \textbf{42}, 114013 (2009).
\bibitem{shapiro1994} J. H. Shapiro and K.-X. Sun, Semiclassical versus quantum behavior in fourth-order interference, J. Opt. Soc. Am. B \textbf{11}, 1130 (1994).
\end{thebibliography}

\end{document}